\documentclass[12pt]{iopart}
\expandafter\let\csname equation*\endcsname\relax
\expandafter\let\csname endequation*\endcsname\relax

\usepackage{amsmath}
\usepackage{amssymb}
\usepackage{graphicx}
\usepackage{braket}
\usepackage{stmaryrd}
\usepackage{hyperref}
\usepackage{color}
\usepackage{dsfont}

\hypersetup{
    colorlinks,
    citecolor=blue,
    linkcolor=blue,
    urlcolor=blue
}

\newcommand{\dblbrace}[1]{\llbracket #1\rrbracket}

\begin{document}

\title[]{Tradeoff Relations in Open Quantum Dynamics via Robertson, Maccone-Pati, and  Robertson-Schr\"odinger  Uncertainty Relations}

\author{Tomohiro Nishiyama}
\ead{htam0ybboh@gmail.com}
\address{Independent Researcher, Tokyo 206-0003, Japan}

\author{Yoshihiko Hasegawa\footnote{Author to whom any correspondence should be addressed}}
\ead{hasegawa@biom.t.u-tokyo.ac.jp}
\address{Department of Information and Communication Engineering, Graduate
School of Information Science and Technology, The University of Tokyo,
Tokyo 113-8656, Japan}

\vspace{10pt}
\begin{abstract}
The Heisenberg uncertainty relation, together with Robertson's generalisation, serves as a fundamental concept in quantum mechanics, showing that noncommutative pairs of observables cannot be measured precisely. 
In this study, we explore the Robertson-type uncertainty relations to demonstrate their effectiveness in establishing a series of thermodynamic uncertainty relations and quantum speed limits in open quantum dynamics.
The derivation utilises a scaled continuous matrix product state representation that maps the time evolution of the quantum continuous measurement to the time evolution of the system and field. 
Specifically, we consider the Maccone-Pati uncertainty relation, a refinement of the Robertson uncertainty relation, to derive thermodynamic uncertainty relations and quantum speed limits. 
These newly derived relations, which use a state orthogonal to the initial state, yield bounds that are tighter than previously known bounds. 
Moreover, we consider the Robertson-Schr\"odinger uncertainty, which extends the Robertson uncertainty relation. 
Our findings not only reinforce the significance of the Robertson-type uncertainty relations, but also expand its applicability in identifying uncertainty relations in open quantum dynamics. 
\end{abstract}

\section{Introduction}

The Heisenberg uncertainty relation \cite{Heisenberg:1927:UR} is a fundamental concept in quantum mechanics, which posits that it is impossible to simultaneously determine both the exact position and the exact momentum of a particle. The more accurately one of these values is known, the less accurate the measurement of the other. 
The Robertson uncertainty relation is a generalisation of the Heisenberg uncertainty relation that can handle general observables other than position or momentum \cite{Robertson:1929:UncRel}.
Let $A$ and $B$ be arbitrary observables and $\ket{\psi}$ be a state vector. 
The Robertson uncertainty relation is 
\begin{align}
    \dblbrace{A}^{2}\dblbrace{B}^{2}\ge\frac{1}{4}|\braket{[A,B]}|^{2},
    \label{eq:robertson_UR_def}
\end{align}
where
$\braket{\bullet}\equiv\braket{\psi|\bullet|\psi}$, $[A,B] \equiv AB-BA$, and $\dblbrace{A}\equiv\sqrt{\braket{A^{2}}-\braket{A}^{2}}$ is the standard deviation of $A$. 
The Robertson uncertainty relation has been used to derive many other uncertainty relations. 
A notable example of this is the quantum speed limit (QSL),
established in 1945 \cite{Mandelstam:1945:QSL} (see Ref.~\cite{Deffner:2017:QSLReview} for a review of QSL).
The QSL can be derived by taking $A=\ket{\psi(0)}\bra{\psi(0)}$ and $B=H$ in Eq.~\eqref{eq:robertson_UR_def},
where $\ket{\psi(0)}$ is the initial state and $H$ is the Hamiltonian. 
Specifically, the Mandelstam-Tamm bound \cite{Mandelstam:1945:QSL} provides a lower bound on the time required for a quantum state to evolve into an orthogonal state. This minimum time $\tau$ satisfies
\begin{align}
    \tau\geq\frac{\pi}{2\dblbrace{H}}.
    \label{eq:MT_bound}
\end{align}
The QSL sets the maximum speed at which a quantum state can evolve and can be understood as a tradeoff between energy and time, highlighting the inherent limitations of quantum systems.
The thermodynamic uncertainty relation (TUR) is a tradeoff in stochastic \cite{Barato:2015:UncRel,Gingrich:2016:TUP,Horowitz:2019:TURReview} or quantum thermodynamics \cite{Erker:2017:QClockTUR,Carollo:2019:QuantumLDP,Hasegawa:2020:QTURPRL}, illustrating that increasing the accuracy of thermodynamic systems can be achieved at the expense of greater thermodynamic resources, such as entropy production and dynamical activity.
Let $\mathcal{A}_\mathrm{cl}(\tau)$ be the time-integrated classical dynamical activity:
\begin{align}
    \mathcal{A}_\mathrm{cl}(\tau)\equiv\int_{0}^{\tau}dt\sum_{\nu,\mu\,(\nu\ne\mu)}P(\mu;t)W_{\nu\mu}.
    \label{eq:Acl_def}
\end{align}
Here, $P(\mu;t)$ is the probability of being the $\mu$th state at time $t$ and $W_{\nu\mu}$ is the transition rate from the $\mu$th state to the $\nu$th state. 
$\mathcal{A}_\mathrm{cl}(\tau)$ in Eq.~\eqref{eq:Acl_def} measures the system's activity by calculating the average number of jumps that occur within the time interval $[0,\tau]$. This provides a quantitative evaluation of how active the system is.
In classical Markov processes, a well-known instance of TURs is \cite{Garrahan:2017:TUR,Terlizzi:2019:KUR}
\begin{align}
    \frac{\dblbrace{\mathcal{C}}^2}{\braket{\mathcal{C}}^{2}}\geq\frac{1}{\mathcal{A}_\mathrm{cl}^\mathrm{ss}(\tau)},
    \label{eq:DAtype_TUR}
\end{align}
where $\mathcal{C}$ is the counting observable counting the number of jumps within the time interval $[0,\tau]$ and 
$\mathcal{A}_\mathrm{cl}^\mathrm{ss}(\tau)$
is $\mathcal{A}_\mathrm{cl}(\tau)$ under the steady-state condition. 
As shown later, a quantum generalisation of Eq.~\eqref{eq:Acl_def}, quantum dynamical activity [cf. Eq.~\eqref{eq:QDA_exact_solution}], plays a central role in tradeoff relations in the present study.

Recently, Ref.~\cite{Hasegawa:2023:BulkBoundaryBoundNC} showed that a quantum TUR can be derived using the Robertson uncertainty relation by mapping a quantum Markov process to the equivalent unitary dynamics through a scaled continuous matrix product state (cMPS) representation. 
These two relations, QSL and TUR, which are derived from the Robertson uncertainty relation, indicate that the Robertson uncertainty relation can be regarded as one of the most significant relations in quantum mechanics. 

In this study, we show that several TURs and QSLs in open quantum dynamics can be derived using the 
Robertson-type uncertainty relations.
The major contribution of the present study over Refs.~\cite{Mandelstam:1945:QSL,Hasegawa:2023:BulkBoundaryBoundNC}, which employed the Robertson uncertainty relation to derive QSL and TUR, is that we apply uncertainty relations other than Robertson's in an open quantum dynamics scenario. 
This is made possible by evaluating the expectation of the induced Hamiltonian (cf. Eqs.~\eqref{eq:Hamiltonian_def} and \eqref{eq:Psi_H_OverlinePsi}),
which requires nontrivial calculations owing to the complicated time-ordering operator. 
Specifically, we apply the Maccone-Pati uncertainty relation \cite{Maccone:2014:UR}, which is an improvement of the Robertson uncertainty relation and takes advantage of a state that is orthogonal to the initial state. 
When we apply the Maccone-Pati uncertainty relation, we can obtain enhanced TURs and QSLs in the open quantum dynamics. Furthermore, the Robertson-Schr{\"o}dinger uncertainty relation is recognised as an refinement of the Robertson uncertainty relation, 
which incorporates not only the commutation relation between two observables but their anti-commutation relation. 
Using the Robertson-Schr\"odinger uncertainty relation, we can derive tighter TURs and QSLs.
Tighter bounds in quantum systems offer several benefits. 
They provide a more precise understanding of quantum mechanics and thermodynamics and enable more accurate predictions. Moreover, improved bounds can result in improved quantum control strategies, potentially leading to faster and more precise quantum operations across various applications.

\section{Methods}

\subsection{Lindblad equation and continuous measurement}

We consider a quantum Markov process described by the Lindblad equation. Let $\rho_S(t)$ be the density operator at time $t$. The Lindblad equation is expressed as $\dot{\rho}_S = \mathcal{L}\rho_S$, where $\mathcal{L}$ denotes the Lindblad superoperator:
\begin{align}
    \mathcal{L}\rho_{S}=-i[H_{S},\rho_{S}]+\sum_{m=1}^{N_{C}}\mathcal{D}\left[L_{m}\right]\rho_{S},
    \label{eq:GKSL_eq_def}
\end{align}
where $H_S$ is the system Hamiltonian, $L_m$ is $m$th jump operator, $N_C$ is the number of jump channels, and $\mathcal{D}[L] \rho_S = L \rho_S L^{\dagger}-\frac{1}{2}\left\{L^{\dagger} L, \rho_S\right\}$. The Lindblad equation is capable of modelling both classical Markov processes and closed quantum systems. By setting $H_S$ to zero, the equation is simplified to describe a classical Markov process. Conversely, closed quantum dynamics can be represented by setting $L_m$ to zero for all $m$. 
Mathematically, any completely positive and trace-preserving (CPTP) map can be described by the Lindblad equation, provided that the dynamics of the system satisfies the semigroup property (also known as the Markovian property). From a physical perspective, the Lindblad equation is applicable only when the interaction between the environment and the principal system is sufficiently weak. This weak coupling approximation relies on three key assumptions: the Born approximation, the Markov approximation, and the rotating wave approximation \cite{Breuer:2002:OpenQuantum}. 

Let $H_{\mathrm{eff}}\equiv H_S-\frac{i}{2}\sum_m L_m^\dagger L_m$ be the effective (non-Hermitian) Hamiltonian. 
The Lindblad equation also has the Kraus representation:
\begin{align}
    \rho_{S}(t+dt)=\sum_{m=0}^{N_{C}}V_{m}(dt)\rho_{S}(t)V_{m}(dt)^{\dagger},
    \label{eq:Kraus_UM_def}
\end{align}
where $V_m$ is the Kraus operator defined by 
$V_{0}(dt)=\mathbb{I}_{S}-idtH_{\mathrm{eff}}$ and $V_{m}(dt)=\sqrt{dt}L_{m}$ ($1 \le m \le N_C$). 
The Kraus operators satisfy the completeness relation $\sum_{m=0}^{N_{C}}V_{m}^{\dagger}(dt)V_{m}(dt)=\mathbb{I}_{S}$. 
By substituting $V_{0}(dt)=\mathbb{I}_{S}-idtH_{\mathrm{eff}}$ and $V_{m}(dt)=\sqrt{dt}L_{m}$ (where $1 \le m \le N_C$) into 
$\rho_{S}(t+dt)=\sum_{m=0}^{N_{C}}V_{m}(dt)\rho_{S}(t)V_{m}(dt)$,
we obtain the Lindblad equation
$d\rho_S = \left(-i[H_{S},\rho_{S}]+\sum_{m=1}^{N_{C}}\mathcal{D}\left[L_{m}\right]\rho_{S}\right)dt + O(dt^2)$, where we have ignored terms of order $dt^2$ and higher.
 When dealing with infinitesimals with respect to time, we usually ignore terms that are quadratic or of higher order.  
From the Stinespring representation, the Kraus representations in Eq.~\eqref{eq:Kraus_UM_def} can be regarded as application of the measurement to the environment at each step,
where $m$ denotes measurement output. 
In this context, the output of \( m=0 \) indicates a ``no jump'' case, which means that there are no jumps in the interval \([t, t+dt]\). In contrast, an output of \( m \geq 1 \) means that the \( m \)th jump occurs within the same interval.
Specifically, if we measure the output $m\ge 0$, the state becomes
$\rho_{S}(t)\to V_{m}(dt)\rho_{S}(t)V_{m}(dt)^{\dagger}/p_{m}$,
where $p_{m}=\mathrm{Tr}_{S}[V_{m}(dt)\rho_{S}(t)V_{m}(dt)^{\dagger}]$ denotes the probability of obtaining $m$. 
Therefore, the dynamics conditioned on the measurement output is identified as a consequence of continuous measurement applied to the environment. 
The record of the output $m$ due to continuous measurement is called a quantum trajectory. For detailed information on continuous measurement, refer to the recent review paper \cite{Landi:2023:CurFlucReview}.

\subsection{Continuous matrix product state}

\begin{figure}
    \centering
    \includegraphics[width=0.85\linewidth]{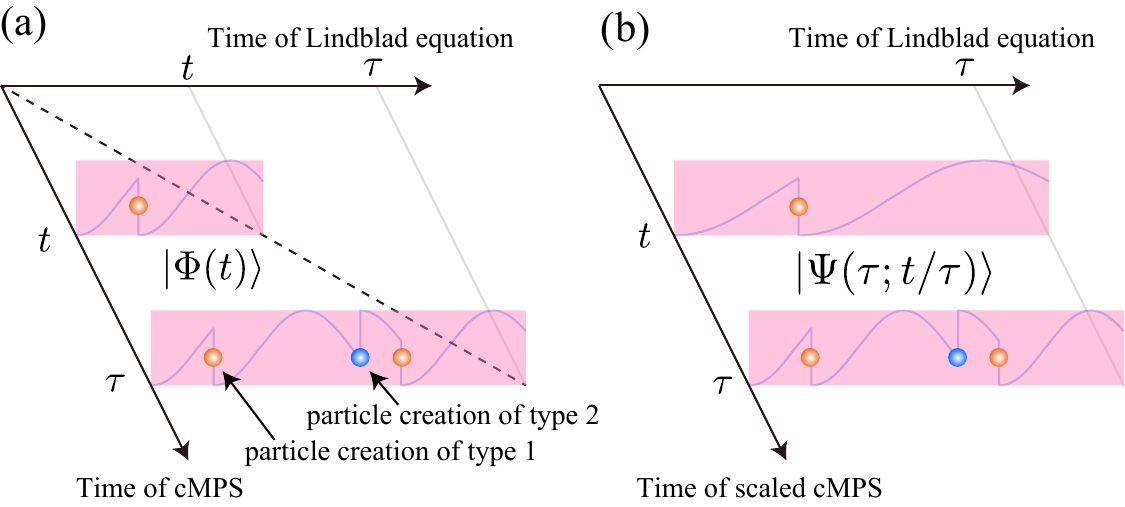}
    \caption{ Intuitive illustration of the cMPS. 
(a) The original cMPS, denoted as $\ket{\Phi(t)}$ and 
(b) scaled cMPS, represented as $\ket{\Psi(\tau;t/\tau)}$. 
In both illustrations, the horizontal axis represents the time evolution of the original Lindblad equation. The vertical axis, moving downward, indicates the time evolution of the cMPS.
The plates indicate the field where the particle creation operator is applied and the lines on the plates denote dynamics conditioned on the measurement record. 
The particle creation of $m$th type is applied on the plates where the $m$th jump occurs.}
    \label{fig:cMPS}
\end{figure}

Having introduced the Lindblad equation, we now review the cMPS representation \cite{Verstraete:2010:cMPS,Osborne:2010:Holography}.
An MPS is a representation of a tensor network in a quantum state. 
cMPS is a generalisation of MPS that can handle continuous coordinates. 
As mentioned above, 
a quantum trajectory is obtained by applying continuous measurement. 
In cMPS, quantum trajectories are recorded onto a quantum field.
Suppose that a quantum trajectory comprises $[(t_1,m_1),(t_2,m_2),\cdots,(t_K,m_K)]$, 
where $m_i \ge 1$ denotes the type of jump that occurs at $t_i$. 
This quantum trajectory can then be recorded in the quantum field using
\begin{align}
\phi_{m_{K}}^{\dagger}\left(t_{K}\right)\cdots\phi_{m_{2}}^{\dagger}\left(t_{2}\right)\phi_{m_{1}}^{\dagger}\left(t_{1}\right)\ket{\mathrm{vac}},
\label{eq:QFT_record}
\end{align}
where $\ket{\mathrm{vac}}$ is the vacuum state and $\phi_m(s)$ is the field operator satisfying the canonical commutation relation $\left[\phi_m(s), \phi_{m^{\prime}}^{\dagger}\left(s^{\prime}\right)\right]=\delta_{m m^{\prime}} \delta\left(s-s^{\prime}\right)$. In Eq.~\eqref{eq:QFT_record},
$\phi_m^{\dagger}(t)$ creates a particle of type $m$ at time $t$.
The time evolution of the principal system, whose dynamics is governed by the Lindblad equation [Eq.~\eqref{eq:GKSL_eq_def}]. The field that records the quantum trajectory can be represented by the following cMPS:
\begin{align}
    \ket{\Phi\left(t\right)}\equiv\mathfrak{U}\left(t\right)\ket{\psi_{S}(0)}\otimes\ket{\mathrm{vac}},
    \label{eq:cMPS_def}
\end{align}
where
$\ket{\psi_S(0)}$ is the initial state of the principal system and $\mathfrak{U}(t)$ is the operator given by
$\mathfrak{U}\left(t\right)\equiv\mathbb{T}\exp[\int_{0}^{t}ds\left(-iH_{S}\otimes\mathbb{I}_{F}+\sum_{m}(L_{m}\otimes\phi_{m}^{\dagger}(s)-L_{m}^{\dagger}\otimes\phi_{m}(s))\right)]$. 
Since the exponent $-iH_{S}\otimes\mathbb{I}_{F}+\sum_{m}(L_{m}\otimes\phi_{m}^{\dagger}(s)-L_{m}^{\dagger}\otimes\phi_{m}(s))$ is skew-Hermitian, it follows that $\mathfrak{U}\left(t\right)$ is unitary. 
As the integration range in $\mathfrak{U}(t)$ depends on $t$, we cannot evaluate the fidelity $\braket{\Phi(t_2)|\Phi(t_1)}$ for $t_1 \ne t_2$. 
Therefore, in Ref.~\cite{Hasegawa:2023:BulkBoundaryBoundNC}, a scaled cMPS representation was introduced:
\begin{align}
    \ket{\Psi\left(\tau;\theta\right)}&\equiv\mathcal{U}\left(\tau,0;\theta\right)\ket{\psi_{S}(0)}\otimes\ket{\mathrm{vac}}\nonumber\\&=\mathcal{V}\left(\tau,0;\theta\right)\ket{\psi_{S}(0)}\otimes\ket{\mathrm{vac}},\label{eq:scaled_cMPS_def}
\end{align}
where
$\theta \equiv t / \tau$ is the scaled time, 
and 
\begin{align}
    \mathcal{U}(s_{2},s_{1};\theta)&\equiv\mathbb{T}\exp\left[\int_{s_{1}}^{s_{2}}ds\left(-i\theta H_{S}\otimes\mathbb{I}_{F}+\sqrt{\theta}\sum_{m}\left(L_{m}\otimes\phi_{m}^{\dagger}(s)-L_{m}^{\dagger}\otimes\phi_{m}(s)\right)\right)\right], \label{eq:def_U}\\
    \mathcal{V}\left(s_{2},s_{1};\theta\right)&\equiv\mathbb{T}\exp\left[\int_{s_{1}}^{s_{2}}ds\left(-i\theta H_{\mathrm{eff}}\otimes\mathbb{I}_{F}+\sqrt{\theta}\sum_{m}L_{m}\otimes\phi_{m}^{\dagger}(s)\right)\right].
    \label{eq:def_V}
\end{align}
In Eq.~\eqref{eq:scaled_cMPS_def}, the second line follows from the canonical
commutation relation and the annihilation of the vacuum by $\phi_m(s)$.
Therefore, $\mathcal{U}$ and $\mathcal{V}$ are different when applied to a general field state, but agree when applied to $\ket{\mathrm{vac}}$. 
We observe that the integration range of $\mathcal{V}(\tau,0;\theta)$ is invariant to changes in $\theta$. 
In the scaled cMPS, the Hamiltonian and jump operators are scaled by $t/\tau$ and $\sqrt{t/\tau}$, respectively.
Therefore, we can calculate the fidelity $\braket{\Psi(\tau;t_2 / \tau) | \Psi(\tau;t_1 / \tau)}$ for $t_1 \ne t_2$. 
Figure~\ref{fig:cMPS} illustrates (a) the original cMPS $\ket{\Phi(t)}$ and (b) the scaled cMPS $\ket{\Psi(\tau;t/\tau)}$. 
The creation operator is applied to the field at the times when the jump occurs. 
In the original cMPS, the jumps in the continuous measurement are recorded on a field defined over the interval $[0, t]$. The scaled cMPS scales the dynamics from the original interval $[0, t]$ to a new interval $[0, \tau]$. As a result, the jump record is now captured in a field defined over $[0, \tau]$. This scaling allows for evaluation of the fidelity.
For details of the scaled cMPS, see Ref.~\cite{Hasegawa:2023:BulkBoundaryBoundNC}.

\subsection{Induced Hamiltonian and dynamical activity}

Because the scaled cMPS representation defined in Eq.~\eqref{eq:def_U} is unitary for the same reason as $\mathfrak{U}\left(t\right)$, the underlying Hamiltonian can be considered. 
For notational simplicity, let $U(t) \equiv \mathcal{U}(\tau,0;\theta = t/\tau)$. 
We can deduce the Hamiltonian as
\begin{align}
    \mathcal{H}(t)&\equiv i\frac{dU(t)}{dt}U^\dagger(t),
    \label{eq:Hamiltonian_def}
\end{align}
which implies $U(t) = \mathbb{T}e^{-i\int_0^t \mathcal{H}(t')dt'}$ ($\mathbb{T}$ denotes the time-ordering operator). 
Hereafter, we call $\mathcal{H}(t)$ the \textit{induced Hamiltonian}. 
Therefore, $\mathcal{H}(t)$ serves as a generator of time evolution in Eq.~\eqref{eq:scaled_cMPS_def}. 
Note that $\mathcal{H}(t)$ is Hermitian because $\frac{dU(t)}{dt}U(t)^\dagger+U(t)\frac{dU(t)^\dagger}{dt}=0$, which can be derived from $U(t)U(t)^\dagger=\mathbb{I}$. 
Then, the scaled cMPS obeys the Schr\"odinger equation:
\begin{align}
    i\frac{d}{dt}\ket{\Psi(t)} =\mathcal{H}(t)\ket{\Psi(t)},
    \label{eq:Schrodinger_eq}
\end{align}
where, for the sake of simplifying the notation, we used $\ket{\Psi(t)} = \ket{\Psi(\tau;t/\tau)}$.
Now, the time evolution of the cMPS can be represented by a closed quantum dynamics, implying that any relation that holds for closed quantum systems should hold for Eq.~\eqref{eq:Schrodinger_eq}. 
The mean and variance of $\mathcal{H}(t)$ are calculated analytically as follows:
\begin{align}
&\bra{\Psi(t)}\mathcal{H}(t)^{2}\ket{\Psi(t)}-\bra{\Psi(t)}\mathcal{H}(t)\ket{\Psi(t)}^{2}=\frac{\mathcal{B}(t)}{4t^{2}}.\label{eq:mathcalH_var}
\end{align}
Here, $\mathcal{J}(t) = \mathcal{B}(t) / t^2$ corresponds to the the quantum Fisher information when $t$ is considered a parameter to be estimated. 
Following Ref.~\cite{Hasegawa:2023:BulkBoundaryBoundNC},
we identify $\mathcal{B}(t) \equiv \mathcal{J}(t)t^2$ as quantum dynamical activity, which is a quantum generalisation of
the classical dynamical activities. 
The quantum dynamic activity plays an important role in quantum TURs and QSLs. 
Recently, we derived an exact representation of $\mathcal{B}(t)$ in Ref.~\cite{Nishiyama:2024:ExactQDAPRE}:
\begin{align}
    \mathcal{B}(\tau)=\mathcal{A}(\tau)+\mathcal{B}_{q}(t),
    \label{eq:QDA_exact_solution}
\end{align}
where
$\mathcal{A}(t)$ and 
$\mathcal{B}(t)$ denote the classical dynamical activity and the quantum contribution, respectively:
\begin{align}
    \mathcal{A}(t)& \equiv\int_{0}^{t}\sum_{m}\mathrm{Tr}_{S}[L_{m}\rho_{S}(s)L_{m}^{\dagger}]ds,\label{eq:classical_DA_def}\\
    \mathcal{B}_{q}(t)&\equiv 8\int_{0}^{t}ds_{1}\int_{0}^{s_{1}}ds_{2}\mathrm{Re}[\mathrm{Tr}_{S}\{H_{\mathrm{eff}}^{\dagger}\check{H}_{S}(s_{1}-s_{2})\rho_{S}(s_{2})\}]-4\left(\int_{0}^{\tau}ds\mathrm{Tr}_{S}\left[H_{S}\rho_{S}(s)\right]\right)^{2}.
    \label{eq:QDA_Bq_def}
\end{align}
Here, 
$\check{H}_S(t) \equiv e^{\mathcal{L}^\dagger t}H_S$ is the Heisenberg interpretation of the Hamiltonian $H_S$
with 
$\mathcal{L}^{\dagger}\mathcal{O}\equiv i\left[H_{S},\mathcal{O}\right]+\sum_{m=1}^{N_{C}}\mathcal{D}^{\dagger}[L_{m}]\mathcal{O}$ being the adjoint superoperator and
$\mathcal{D}^\dagger$ being the adjoint dissipator $\mathcal{D}^{\dagger}[L]\mathcal{O}\equiv L^{\dagger}\mathcal{O} L-\frac{1}{2}\left\{ L^{\dagger}L,\mathcal{O}\right\} $. 
The contribution of $\mathcal{B}_q(t)$ enables a quantum system to perform more accurately and rapidly. 
Note that an intermediate expression of the quantum dynamical activity was derived analytically in Ref.~\cite{Nakajima:2023:SLD}. 
The classical limit demonstrates the equivalence between $\mathcal{A}_\mathrm{cl}(\tau)$ in Eq.~\eqref{eq:Acl_def} and $\mathcal{A}(\tau)$ in Eq.~\eqref{eq:classical_DA_def}. This can be verified by two steps. First, we assign $L_m = L_{\nu\mu} = \sqrt{W_{\nu \mu}}\ket{\nu}\bra{\mu}$, where $\ket{\mu}$ corresponds to the $\mu$th state in a classical Markov process. Second, we consider a density operator $\rho_S(t)$ whose diagonal elements correspond to the probability distribution $P(\mu;t)$.

\section{Results}
This section demonstrates the derivation of TURs and QSLs using the Robertson, Maccone-Pati, and Robertson-Schrödinger uncertainty relations.
The relations derived in this study are summarised in Table~\ref{tab:UR_relation}.

\begin{table}
\centering
\begin{tabular}{|c|c|c|c|}
\hline 
Uncertainty relation & $\begin{array}{c}
\scriptstyle A=\mathbb{I}_{S}\otimes\mathcal{C}_{F}\\
\scriptstyle B=\mathcal{H}(t)
\end{array}$ & $\begin{array}{c}
\scriptstyle A=\mathcal{C}_{S}\otimes \mathbb{I}_F\\
\scriptstyle B=\mathcal{H}(t)
\end{array}$ & $\begin{array}{c}
\scriptstyle A=\ket{\Psi(0)}\bra{\Psi(0)}\\
\scriptstyle B=\mathcal{H}(t)
\end{array}$\\
\hline 
\hline 
$\scriptstyle \dblbrace{A}^{2}\dblbrace{B}^{2}\ge\frac{1}{4}|\braket{[A,B]}|^{2}$  \cite{Heisenberg:1927:UR,Robertson:1929:UncRel} & \footnotesize Eq.~\eqref{eq:QTUR_from_Robertson} & -- & \footnotesize Eq.~\eqref{eq:QSL_open_quantum}\\
\hline 
$\scriptstyle \dblbrace{A}^{2}+\dblbrace{B}^{2}\geq\pm i\braket{[A,B]}+\left|\bra{\psi}\left(A\pm iB\right)\ket{\overline{\psi}}\right|^{2}$ \cite{Maccone:2014:UR} & \footnotesize Eq.~\eqref{eq:main_result_1} & \footnotesize Eq.~\eqref{eq:main_result_1} & -- \\
\hline 
$\scriptstyle \dblbrace{A}\dblbrace{B}\geq\pm\frac{\frac{i}{2}\langle[A,B]\rangle}{1-\frac{1}{2}\left|\Braket{\psi\left|\frac{A}{\dblbrace{A}}\pm i\frac{B}{\dblbrace{B}}\right|\overline{\psi}}\right|^{2}}$ \cite{Maccone:2014:UR} & \footnotesize Eq.~\eqref{eq:main_result_2} & \footnotesize Eq.~\eqref{eq:main_result_2} & \footnotesize Eq.~\eqref{eq:MacconePati_QSL}\\
\hline 
$\scriptstyle \dblbrace{A}^{2}\dblbrace{B}^{2}\geq\left(\frac{1}{2}\braket{\{A,B\}}-\braket{A}\braket{B}\right)^{2}+\frac{1}{4}\left|\braket{[A,B]}\right|^{2}$ \cite{Schrodinger:1930:UR} & \footnotesize Eq.~\eqref{eq:new_rsr_field_main} & \footnotesize Eq.~\eqref{eq:new_rsr_main} & \footnotesize Eq.~\eqref{eq:RSUR_QSL}\\
\hline 
\end{tabular}

\caption{Interconnections between uncertainty relations, observables $A$ and $B$, and their corresponding TURs and QSLs. 
In the leftmost column, $\ket{\psi}$ is a general state vector and $\ket{\overline{\psi}}$ is a state vector orthogonal to $\ket{\psi}$, i.e., $\braket{\psi|\overline{\psi}} = 0$. 
$\dblbrace{A}$ denotes the standard derivation of the observable $A$. 
The leftmost column lists four types of inequalities. In descending order, they are: Robertson, Maccone-Pati (sum type), Maccone-Pati (product type), and Robertson-Schrodinger uncertainty relations.
The results of the second and the fourth columns correspond to TUR and QSL, respectively. 
\label{tab:UR_relation}}

\end{table}

\subsection{TUR from Robertson uncertainty relation\label{sec:TUR_RUR_main}}

Consider the Robertson uncertainty relation [Eq.~\eqref{eq:robertson_UR_def}] in Eq.~\eqref{eq:Schrodinger_eq}. 
Suppose $A = \mathbb{I}_S \otimes \mathcal{C}_F$ and $B = \mathcal{H}(t)$,
where $\mathcal{C}_F$ denotes a measurement operator in the field.
For example, for $\mathcal{C}_F$, we consider the total number operators that count the number of jump events within $[0,\tau]$. 
As the variance of $\mathcal{H}(t)$ is given by Eq.~\eqref{eq:mathcalH_var}, the following relation can be derived from Eq.~\eqref{eq:robertson_UR_def} \cite{Hasegawa:2023:BulkBoundaryBoundNC}:
\begin{align}
    \frac{\dblbrace{\mathcal{C}_{F}}_{F}(t)^{2}}{t^{2}\left(\partial_{t}\braket{\mathcal{C}_{F}}_{F}(t)\right)^{2}}\geq\frac{1}{\mathcal{B}(t)},
    \label{eq:QTUR_from_Robertson}
\end{align}
where $\braket{\bullet}_{Z}(t)\equiv\mathrm{Tr}_{Z}[\bullet\rho_{Z}(t)]$ is the mean and $\dblbrace{\bullet}_{Z}(t)\equiv\sqrt{\braket{\bullet^{2}}_{Z}(t)-\braket{\bullet}_{Z}(t)^{2}}$ is the standard deviation. 
 The subscript $Z$ is either $S$ or $F$, and $\rho_{F}(t)=\mathrm{Tr}_S[\ket{\Psi(t)}\bra{\Psi(t)}]$, $\rho_{S}(t)=\mathrm{Tr}_F[\ket{\Psi(t)}\bra{\Psi(t)}]$.
The details of the derivation are shown in \ref{sec:TUR_QSL_by_RUR}.

\subsection{QSL from Robertson uncertainty relation\label{sec:QSL_RUR_main}}

Next, we derive a QSL for open quantum dynamics using the Robertson uncertainty relation. 
Although the basic approach follows the one described in Ref.~\cite{Mandelstam:1945:QSL}, obtaining a QSL using the scaled cMPS via the Robertson uncertainty relation has not yet been reported. 
Therefore, to ensure the integrity of this study, we demonstrate that the QSL can be derived through the Robertson uncertainty relation.
Taking $A = \ket{\Psi(0)}\bra{\Psi(0)}$ and $B = \mathcal{H}(t)$,
which is an analogue of the Mandelstam-Tamm bound,
we obtain the QSL for open quantum systems:
\begin{align}
    \frac{1}{2}\int_0^t ds \frac{\sqrt{\mathcal{B}(s)}}{s} \geq \mathcal{L}_D(\ket{\Psi(t)}, \ket{\Psi(0)})\geq  \mathcal{L}_D(\rho_S(t), \rho_S(0)),
    \label{eq:QSL_open_quantum}
\end{align}
which was derived in Ref.~\cite{Hasegawa:2023:BulkBoundaryBoundNC} using a different approach based on
the geometric QSL. 
Here, $\mathcal{L}_D$ is the Bures angle, defined as follows:
\begin{align}
    \mathcal{L}_D(\rho_1, \rho_2)&\equiv \mathrm{arccos}\left[\sqrt{\mathrm{Fid}(\rho_1, \rho_2)}\right],
    \label{eq:Bures_angle_def}
\end{align}
where $\mathrm{Fid}(\rho_1, \rho_2)$ is the quantum fidelity $\mathrm{Fid}(\rho_1, \rho_2)\equiv \left(\mathrm{Tr}\sqrt{\sqrt{\rho_1}\rho_2\sqrt{\rho_1}}\right)^2$. 
The details of the derivation are shown in \ref{sec:TUR_QSL_by_RUR}. 

\subsection{TUR from Maccone-Pati uncertainty relation\label{sec:TUR_MPUR_main}}

Thus far, we demonstrated that existing uncertainty relations can be derived from the Robertson uncertainty relation. Given that Eq.~\eqref{eq:Schrodinger_eq} represents a closed quantum dynamic, different uncertainty relations can be employed.
We consider the Maccone-Pati uncertainty relation \cite{Maccone:2014:UR},
which is a tighter version of the Robertson uncertainty relation,
to derive TURs and QSLs that are tighter than those previously identified.
The Maccone-Pati uncertainty relation is a refinement of Robertson's given by
\begin{align}
    \dblbrace{A}^2+\dblbrace{B}^2\geq \pm i\braket{[A,B]}+\left|\bra{\psi}\left(A\pm iB\right)\ket{\overline{\psi}}\right|^2,
    \label{eq:Maccone_Pati_def}
\end{align}
where $\ket{\overline{\psi}}$ is an arbitrary state vector orthogonal to $\ket{\psi}$, that is, $\braket{\psi|\overline{\psi}} = 0$. 
In Eq.~\eqref{eq:Maccone_Pati_def}, we should choose the sign (plus or minus) so that the first term on the right-hand side of Eq.~\eqref{eq:Maccone_Pati_def} becomes positive (the inequality itself is valid even if the other sign is used). 
Reference~\cite{Maccone:2014:UR} derived the inequality using a similar approach:
\begin{align}
    \dblbrace{A}\dblbrace{B}\geq\pm\frac{\frac{i}{2}\langle[A,B]\rangle}{1-\frac{1}{2}\left|\Braket{\psi\left|\frac{A}{\dblbrace{A}}\pm i\frac{B}{\dblbrace{B}}\right|\overline{\psi}}\right|^{2}}.
    \label{eq:Maccone_Pati2_def}
\end{align}
We can derive Eq.~\eqref{eq:Maccone_Pati2_def} from Eq.~\eqref{eq:Maccone_Pati_def} by substituting $A / \dblbrace{A}$ and $B / \dblbrace{B}$ into $A$ and $B$ in Eq.~\eqref{eq:Maccone_Pati_def}. 
By excluding the second term in the denominator, the right-hand side of Eq.~\eqref{eq:Maccone_Pati2_def} simplifies to the Robertson uncertainty relation. 
Again, the sign of Eq.~\eqref{eq:Maccone_Pati2_def} should be such that the right-hand side becomes positive. 
Equation~\eqref{eq:Maccone_Pati2_def} is either as tight as or tighter than the Robertson uncertainty relation.
The original Robertson uncertainty relation utilises information only from the state $\ket{\psi}$, while the Maccone-Pati uncertainty relation incorporates additional information obtained from its orthogonal state $\ket{\overline{\psi}}$.
We apply Eq.~\eqref{eq:Maccone_Pati_def} to the scaled cMPS representation [Eq.~\eqref{eq:scaled_cMPS_def}] to derive TURs and QSLs. 
First, we define the state vector in the primary system $\ket{\overline{\psi}_S}$, which is orthogonal to $\ket{\psi_S}$, $\braket{\psi_S|\overline{\psi}_S} = 0$.
If the orthogonality is satisfied, we can arbitrarily choose $\ket{\overline{\psi}_S}$. 
Subsequently, we define its associated state as
\begin{align}
    \ket{\overline{\Psi}(t)}&\equiv U(t)\ket{\overline{\psi}_S}\otimes\ket{\mathrm{vac}},
    \label{eq:Psi_overline_def}
\end{align}
which is the scaled cMPS corresponding to $\ket{\overline{\psi}_S}$ and is orthogonal to $\ket{\Psi(t)}$, $\braket{\overline{\Psi}(t) | \Psi(t)} = 0$. 
Let $\chi_F \equiv \mathrm{Tr}_S[\ket{\overline{\Psi}(t)}\bra{\Psi(t)}]$ and 
$\chi_S \equiv \mathrm{Tr}_F[\ket{\overline{\Psi}(t)}\bra{\Psi(t)}]$. 
Having defined the orthogonal state $\ket{\overline{\Psi}(t)}$ in the scaled cMPS representation, we can apply Eqs.~\eqref{eq:Maccone_Pati_def} and \eqref{eq:Maccone_Pati2_def} to $\ket{\Psi(t)}$ and $\ket{\overline{\Psi}(t)}$. 
As shown in Eqs.~\eqref{eq:Maccone_Pati_def} and \eqref{eq:Maccone_Pati2_def}, we need to evaluate the quantity $\braket{\psi |\bullet|\overline{\psi}}$ in the Maccone-Pati uncertainty relation. 
We evaluate the following quantities with respect to the induced Hamiltonian $\mathcal{H}(t)$:
\begin{align}
&\braket{\Psi(t)|\mathcal{H}|\Psi(t)}=i\braket{\Psi(t)|\partial_{t}\Psi(t)}\nonumber\\&\phantom{\braket{\Psi(t)|\mathcal{H}|\Psi(t)}}=\frac{1}{t}\int_{0}^{t}ds\mathrm{Tr}_{S}\left[H_{S}\rho_{S}(s)\right],\label{eq:mathcalH_mean}\\&\bra{\Psi(t)}\mathcal{H}(t)\ket{\overline{\Psi}(t)}=\frac{1}{t}\int_{0}^{t}ds\mathrm{Tr}_{S}\left[H_{S}\chi_{S}(s)\right].\label{eq:Psi_H_OverlinePsi}
\end{align}
Notably, evaluation using Eqs.~\eqref{eq:mathcalH_mean} and \eqref{eq:Psi_H_OverlinePsi} is nontrivial (see
\ref{sec:MacconePatiUR} and \ref{sec:RSUR_TUR_system}).
This is because the time derivative of $\ket{\Psi(t)}$ should be performed in the presence of the time-ordering operator $\mathbb{T}$ (see the definition $\ket{\Psi(t)} = \ket{\Psi(\tau;\theta = t/ \tau)}$ in Eq.~\eqref{eq:scaled_cMPS_def}). 
This evaluation makes it possible to apply uncertainty relations other than Robertson's to the scaled cMPS representation. 
To derive the quantum TUR [Eq.~\eqref{eq:QTUR_from_Robertson}],
we have employed $A = \mathbb{I}_S \otimes \mathcal{C}_F$,
where $\mathcal{C}_F$ denotes an operator in the field. 
For Eqs.~\eqref{eq:Maccone_Pati_def} and \eqref{eq:Maccone_Pati2_def}, we have two options:
\begin{align}
    A &= \mathcal{C}_S \otimes \mathbb{I}_F,\label{eq:system_observable}\\
    A &= \mathbb{I}_S \otimes \mathcal{C}_F,\label{eq:field_observable}
\end{align}
where $\mathcal{C}_S$ is an observable in the system. 
$\braket{\psi |\bullet|\overline{\psi}}$ term corresponding to $\mathcal{C}_S$ and $\mathcal{C}_F$ is evaluated as follows:
\begin{align}
    \bra{\Psi(t)}\mathbb{I}_{S}\otimes\mathcal{C}_{F}\ket{\overline{\Psi}(t)}&=\mathrm{Tr}_{F}[\mathcal{C}_{F}\chi_{F}(t)],\label{eq:F_braket_main}\\\bra{\Psi(t)}\mathcal{C}_{S}\otimes\mathbb{I}_{F}\ket{\overline{\Psi}(t)}&=\mathrm{Tr}_{S}[\mathcal{C}_{S}\chi_{S}(t)].\label{eq:S_braket_main}
\end{align}
Substituting Eqs.~\eqref{eq:mathcalH_var}, \eqref{eq:Psi_H_OverlinePsi}, \eqref{eq:F_braket_main}, and \eqref{eq:S_braket_main} into Eq.~\eqref{eq:Maccone_Pati_def}, we obtain the following inequality:
\begin{align}
\frac{\mathcal{B}(t)}{4}+t^{2}\llbracket\mathcal{C}_{Z}\rrbracket_{Z}^{2}\geq\pm t^{2}\partial_{t}\langle\mathcal{C}_{Z}\rangle_{Z}+\left|\int_{0}^{t}ds\mathrm{Tr}_{S}\left[H_{S}\chi_{S}(s)\right]\pm it\mathrm{Tr}_{Z}[\mathcal{C}_{Z}\chi_{Z}(t)]\right|^{2}.
    \label{eq:main_result_1}
\end{align}
We dropped $t$ from $\braket{\bullet}_Z$ and $\dblbrace{\bullet}_Z$ for notational simplicity. 
Similarly, instead of using Eq.~\eqref{eq:Maccone_Pati_def}, 
Eq.~\eqref{eq:Maccone_Pati2_def} yields the following inequality:
\begin{align}
\sqrt{\mathcal{B}(t)}\dblbrace{\mathcal{C}_{Z}}_{Z}\geq\frac{\pm t\partial_{t}\braket{\mathcal{C}_{Z}}_{Z}}{1-\frac{1}{2}\left|\frac{2\int_{0}^{t}ds\mathrm{Tr}_{S}\left[H_{S}\chi_{S}(s)\right]}{\sqrt{\mathcal{B}(t)}}\pm i\frac{\mathrm{Tr}_{Z}[\mathcal{C}_{Z}\chi_{Z}(t)]}{\dblbrace{\mathcal{C}_{Z}}_{Z}}\right|^{2}}.
    \label{eq:main_result_2}
\end{align}
In Eqs.~\eqref{eq:main_result_1} and \eqref{eq:main_result_2}, $Z$ is either $S$ or $F$. 
When we ignore the second term in the denominator of Eq.~\eqref{eq:main_result_2}, which is always negative, the relation reduces to quantum TUR when $Z= F$. 
Therefore, when $Z= F$, $\dblbrace{C_Z}_Z = \dblbrace{C_F}_F$ and Eq.~\eqref{eq:main_result_2} can be regarded as a refinement of the quantum TUR given by Eq.~\eqref{eq:QTUR_from_Robertson}. 
While Eq.~\eqref{eq:main_result_2} represents the uncertainty between $\mathcal{B}(t)$ and $\dblbrace{\mathcal{C_Z}}_Z$ by their product,
Eq.~\eqref{eq:main_result_1} does the same using their sum. 
Such TUR-type uncertainty relations, represented by the sum of terms, have not been previously derived. 
$\chi_S(t)$ is the off-diagonal element in the basis of $\ket{\psi_S}$ and $\ket{\overline{\psi}_S}$,
which characterises the quantum coherence. 
Moreover, $\mathrm{Tr}[H_S \chi_S(s)]$ represents the nondiagonal elements with respect to the basis at $s$. 
The off-diagonal elements of the Hamiltonian matrix represent the transition amplitudes or interactions between the different states through the Hamiltonian.

\subsection{QSL from Maccone-Pati uncertainty relation\label{sec:QSL_MPUR_main}}

The original QSL \cite{Mandelstam:1945:QSL} was derived by the Robertson uncertainty. Following a similar approach, a recent study used the Maccone-Pati uncertainty relation to derive another version of the QSL \cite{Thakurai:2022:StrongerQSL}. In this study, we use the Maccone-Pati uncertainty relation to derive a QSL applicable to open quantum dynamics.
Recall that the following relation holds (see 
 \ref{Maccone_QSL}  for details): 
\begin{align}
    \braket{\Psi(t)|\Psi(0)}&=\braket{\psi_{S}|e^{iH_{\mathrm{eff}}^{\dagger}t}|\psi_{S}}=\mathrm{Tr}_{S}[e^{iH_{\mathrm{eff}}^\dagger t}\rho_{S}(0)]\equiv\gamma(t),\label{eq:Psit_Psi0}\\\braket{\Psi(0)|\overline{\Psi}(t)}&=\braket{\psi_{S}|e^{-iH_{\mathrm{eff}}t}|\overline{\psi}_{S}}=\mathrm{Tr}_{S}[e^{-iH_{\mathrm{eff}}t}\chi_{S}(0)].
    \label{eq:Psi0_OverlinePsit}
\end{align}
Substituting $A / \dblbrace{A}$ and $\mathcal{H} / \dblbrace{\mathcal{H}}$ into $A$ and $B$ in Eq.~\eqref{eq:Maccone_Pati_def} and using Eqs.~\eqref{eq:Psi_H_OverlinePsi}, \eqref{eq:Psit_Psi0}, and \eqref{eq:Psi0_OverlinePsit}, we obtain the QSL:
\begin{align}
    \frac{1}{2}\int_{0}^{\tau}dt(1-R(t))\frac{\sqrt{\mathcal{B}(t)}}{t}\geq\mathcal{L}_{D}(\rho_{S}(\tau),\rho_{S}(0)),
    \label{eq:MacconePati_QSL}
\end{align}
where
\begin{align}
    R(t)=\frac{1}{2}\left|\frac{\gamma(t)\mathrm{Tr}_{S}[e^{-iH_{\mathrm{eff}}t}\chi_{S}(0)]}{|\gamma(t)|\sqrt{(1-|\gamma(t)|^{2})}}+i\frac{2\int_{0}^{t}ds\mathrm{Tr}_{S}\left[H_{S}\chi_{S}(s)\right]}{\sqrt{\mathcal{B}(t)}}\right|^{2},
    \label{eq:Rt_def}
\end{align}
and we select the plus sign in Eq.~\eqref{eq:Maccone_Pati_def}. As $1\geq 1-R(t)$ is satisfied, the QSL given by Eq.~\eqref{eq:MacconePati_QSL} is tighter than Eq.~\eqref{eq:QSL_open_quantum}. 
Again, the QSL of Eq.~\eqref{eq:MacconePati_QSL} uses the coherence term $\mathrm{Tr}_S[H_S \chi_S(t)]$. 

\subsection{TUR from Robertson-Schr\"odinger uncertainty relation\label{sec:TUR_RSUR_main}}
So far we have investigated the implications of using the Robertson and Maccone-Pati uncertainty relations to derive TURs and QSLs for open quantum dynamics. 
We next consider the Robertson-Schr\"odinger uncertainty relation to derive TURs and QSLs.
Consider the Robertson-Schr\"odinger uncertainty relation \cite{Schrodinger:1930:UR}:
\begin{align}
    \dblbrace{A}^{2}\dblbrace{B}^{2}\geq\left(\frac{1}{2}\braket{\{A,B\}}-\braket{A}\braket{B}\right)^{2}+\frac{1}{4}\left|\braket{[A,B]}\right|^{2},
    \label{eq:Robertson_schrodinger_def}
\end{align}
where $\{A,B\} \equiv AB + BA$ is the anti-commutator. 
Equation~\eqref{eq:Robertson_schrodinger_def} uses the anticommutator of the two observables to strengthen the Robertson uncertainty relation. 
Removal of the first term from the right-hand side of Eq.~\eqref{eq:Robertson_schrodinger_def} reproduces Eq.~\eqref{eq:robertson_UR_def}. 
Following the procedures employed in deriving TURs using the Maccone-Pati uncertainty relation, we can derive TURs using Eq.~\eqref{eq:Robertson_schrodinger_def}. 

In the TURs based on Maccone-Pati uncertainty relations, we have considered measurement operators in the system and in the field [Eqs.~\eqref{eq:system_observable} and \eqref{eq:field_observable}]. 
In a similar way, we consider measurement operators in the system and in the field. 
We must evaluate the anti-commutator term in Eq.~\eqref{eq:Robertson_schrodinger_def}, which can be calculated as
\begin{align}
    \bra{\Psi(t)}\{\mathcal{C}_S,\mathcal{H}(t)\}\ket{\Psi(t)}=\frac{2}{t}\mathrm{Re}\int_0^{t}ds  \mathrm{Tr}_{S}\left[\check{\mathcal{C}}_S(t-s)H_{\mathrm{eff}}\rho_S(s)\right],
    \label{eq:anticom_main}
\end{align}
where $\check{\mathcal{C}_S}(t)\equiv e^{\mathcal{L}^\dagger t}C_S$ is the Heisenberg interpretation of $\mathcal{C}_S$~\cite{Nishiyama:2024:ExactQDAPRE}.
A derivation of Eq.~\eqref{eq:anticom_main} is detailed in~\ref{sec:RSUR_TUR_system}.
By substituting $\mathcal{C}_{S}=\mathbb{I}_S$ into Eq.~\eqref{eq:anticom_main}, we obtain Eq.~\eqref{eq:mathcalH_mean}.
Setting $A=\mathbb{I}_{F}\otimes \mathcal{C}_S$ and $B=\mathcal{H}(t)$ in Eq.~\eqref{eq:Robertson_schrodinger_def}, and combining 
Eqs.~\eqref{eq:mathcalH_mean} and ~\eqref{eq:anticom_main}, 
we obtain
\begin{align}
\mathcal{B}(t)\llbracket\mathcal{C}_S\rrbracket_S^2\geq 4\left(\mathrm{Re}\int_0^{t}ds  \mathrm{Tr}_{S}\left[\check{\mathcal{C}}_S(t-s)  H_{\mathrm{eff}}\rho_S(s)\right]-\langle\mathcal{C}_S\rangle_S \int_0^{t}ds  \langle H_{S}\rangle_S(s)\right)^2+t^2\left(\partial_t \langle \mathcal{C}_S\rangle_S\right)^2.
\label{eq:new_rsr_main}
\end{align}
Equation~\eqref{eq:new_rsr_main} is a TUR for the system operator obtained by the Robertson-Schr\"odinger equation. 

The field operator case can be obtained in a similar way. 
The anti-commutator term is obtained as (see~\ref{sec:RSUR_TUR_field})
\begin{align}
   &\bra{\Psi(t)}\{\mathcal{C}_{F},\mathcal{H}(t)\}\ket{\Psi(t)}\nonumber\\&=\frac{2}{t}\int_{0}^{t}ds\bra{\Psi(s)}\mathcal{C}_{F}H_{S}\ket{\Psi(s)}+\frac{1}{t}\int_{0}^{t}dss\sum_{m}\alpha_{m}\bra{\Psi(s)}\{L_{m}^{\dagger}L_{m},\mathcal{H}(s)\}\ket{\Psi(s)}\nonumber\\&=\frac{2}{t}\int_{0}^{t}ds\bra{\Psi(s)}\mathcal{C}_{F}H_{S}\ket{\Psi(s)}+\frac{2}{t}\mathrm{Re}\int_{0}^{t}ds_{1}\int_{0}^{s_{1}}ds_{2}\mathrm{Tr}_{S}\left[\check{\mathfrak{L}}(s_{1}-s_{2})H_{\mathrm{eff}}\rho_{S}(s_{2})\right],
\end{align}
where $\mathfrak{L}\equiv \sum_m \alpha_m L_m^\dagger L_m$.
By combining Eqs.~\eqref{eq:mathcalH_mean} and \eqref{eq:Robertson_schrodinger_def} with this equation, we obtain
\begin{align}
\mathcal{B}(t)\llbracket\mathcal{C}_{F}\rrbracket_{F}^{2}&\geq4\Bigl(\int_{0}^{t}ds\bra{\Psi(s)}\mathcal{C}_{F}H_{S}\ket{\Psi(s)}-\langle\mathcal{C}_{F}\rangle_{F}\int_{0}^{t}ds\langle H_{S}\rangle_{S}(s)\nonumber\\&+\mathrm{Re}\int_{0}^{t}ds_{1}\int_{0}^{s_{1}}ds_{2}\mathrm{Tr}_{S}\left[\check{\mathfrak{L}}(s_{1}-s_{2})H_{\mathrm{eff}}\rho_{S}(s_{2})\right]\Bigr)^{2}+t^{2}\left(\partial_{t}\langle\mathcal{C}_{F}\rangle_{F}\right)^{2}.
    \label{eq:new_rsr_field_main}
\end{align}
The first term on the right side of Eq.~\eqref{eq:new_rsr_field_main} originates from the anti-commutation component in the Robertson-Schr\"odinger equation. Given that this term is nonnegative, Eq.~\eqref{eq:new_rsr_field_main} is either as tight as or tighter than the TUR of Eq.~\eqref{eq:QTUR_from_Robertson}.

\subsection{QSL from Robertson-Schr\"odinger uncertainty relation\label{sec:QSL_RSUR_main}}

A QSL for the Robertson-Schr\"odinger uncertainty relation can be obtained by setting $A=\ket{\Psi(0)}\bra{\Psi(0)}$ and $B=\mathcal{H}(t)$ in Eq.~\eqref{eq:Robertson_schrodinger_def}. The main difference from the Maccone-Pati uncertainty relation is the term $\braket{\{A,B\}}$, which can be evaluated as
\begin{align}
    \braket{\{A,B\}}=2\mathrm{Re}\left(\gamma(t)\beta(t)\right),
\end{align}
where $\beta(t)\equiv\mathrm{Tr}_{S}[H_{\mathrm{eff}}\exp(-iH_{\mathrm{eff}}t)\rho_{S}(0)]$. 
Then, we obtain (see \ref{sec:RSUR_QSL})
\begin{align}
    \int_0^\tau dt\sqrt{\frac{\mathcal{B}(t)}{4t^2}-S(t)}\geq \mathcal{L}_D(\rho_S(\tau), \rho_S(0)),
    \label{eq:RSUR_QSL}
\end{align}
where
\begin{align}
     S(t)\equiv \frac{1}{|\gamma(t)|^2(1-|\gamma(t)|^{2})}\left(\mathrm{Re}\left(\gamma(t)\beta(t)\right)-\frac{|\gamma(t)|^2}{t}\int_0^{t}ds  \mathrm{Tr}_{S}\left[H_S\rho_S(s)\right]\right)^2\geq 0.
\end{align}
Since $S(t)$ is always non-negative, the QSL presented in Eq.~\eqref{eq:RSUR_QSL} provides a bound that is at least as tight as, and potentially tighter than, the QSL described in Eq.~\eqref{eq:QSL_open_quantum}.

\section{Discussion}
The Lindblad equation describes open quantum systems, and it can represent closed quantum systems and classical Markov chains as particular cases.
So far, TURs and QSLs have been derived using the Robertson, Maccone-Pati, and Robertson-Schr\"odinger uncertainty relations. 
The Maccone-Pati and Robertson-Schr\"odinger uncertainty relations provide TURs and QSLs that incorporate additional information not accounted for by the Robertson uncertainty relation alone. Specifically, the Maccone-Pati uncertainty relation derives TURs and QSLs using a state that is orthogonal to the initial state. The Robertson-Schr\"odinger uncertainty relation provides TURs and QSLs that are based on the correlation between the induced Hamiltonian, denoted as $\mathcal{H}$, and the measurement operators. 
We can see that the quantum dynamical activity $\mathcal{B}(\tau)$ appears in all these equations. In other words, it suggests that understanding how $\mathcal{B}(\tau)$ behaves is important in all trade-off relations derived here.
Let us discuss two limiting cases of $\mathcal{B}(\tau)$, namely the closed quantum limit and the classical limit. 
In the closed quantum limit, where all Lindblad operators $L_m$ are zero, $\mathcal{B}(\tau)$ takes the form:
\begin{align}
    {\mathcal{B}}(\tau) = 4\tau^{2}\left(\mathrm{Tr}_{S}\left[H_{S}^{2}\rho_{S}\right]-\mathrm{Tr}_{S}\left[H_{S}\rho_{S}\right]^{2}\right)=4\tau^{2}\dblbrace{H_{S}}^{2},
    \label{eq:Bt_quantum_limit}
\end{align}
which represents the variance of $H_S$ multiplied by $4\tau^2$. When we substitute Eq.~\eqref{eq:Bt_quantum_limit} into Eq.~\eqref{eq:QSL_open_quantum}, we recover the Mandelstam-Tamm bound \cite{Mandelstam:1945:QSL}, as shown in Eq.~\eqref{eq:MT_bound}.
In contrast, in the classical limit, where the system Hamiltonian $H_S$ is zero, $\mathcal{B}(\tau)$ simplifies to
\begin{align}
    \mathcal{B}(\tau) = \mathcal{A}(\tau) = \int_{0}^{t}\sum_{m}\mathrm{Tr}_{S}[L_{m}\rho_{S}(s)L_{m}^{\dagger}]ds.
    \label{eq:Bt_classical_limit}
\end{align}
Here, $\mathcal{A}(\tau)$ is the classical dynamical activity [Eq.~\eqref{eq:classical_DA_def}]. 
Comparing Eqs.~\eqref{eq:Bt_quantum_limit} and \eqref{eq:Bt_classical_limit}, we see that, in the case of a closed quantum system, $\mathcal{B}(\tau)$ behaves as $O(\tau^2)$, whereas for a classical Markov chain, it is $O(\tau)$. 
This indicates that in a closed quantum system, the state can change more significantly due to coherent dynamics compared to when there is interaction with the environment.
For an open quantum system, it becomes larger than $O(\tau)$ up to a certain point in time, and in the limit as $\tau \to \infty$, it approaches $O(\tau)$ similarly to the classical case. 

In this study, we derived TURs and QSLs from Robertson-like uncertainty relations, which are tighter than previously known relations. These improved bounds in quantum systems offer several benefits. Notably, they can lead to enhanced quantum control strategies, potentially resulting in faster and more precise quantum operations. The practical applications of these tighter bounds in optimising quantum devices and protocols would be the focus of our future studies.

\appendix

\section{TUR and QSL from Robertson uncertainty relation\label{sec:TUR_QSL_by_RUR}}
We show that we can derive TURs and QSLs from the Robertson uncertainty relation in this section.
 Letting $\theta \equiv t/\tau$, for  notational convenience, we define
\begin{align}
    \label{def:psi_t}
    \ket{\Psi(t)}&=\ket{\Psi\left(\tau;\theta\right)}.
\end{align}
Since the odd order terms of $\phi_m(s)$ and $\phi_m(s)^\dagger$ vanish in $\rho_S(\tau; \theta)=\mathrm{Tr}_F\left[\ket{\Psi\left(\tau;\theta\right)}\bra{\Psi\left(\tau;\theta\right)}\right]$, the density matrix $\rho_S(\tau; \theta)$ can be expanded in terms proportional to $\theta ds$, that is, the system $\rho_S(\tau; \theta)$ evolves $\theta$ times as fast as the dynamics when $\theta=1$. Therefore, we obtain 
\begin{align}
    \rho_S(\tau; \theta)=\rho_S(t; 1)=\mathrm{Tr}_F\left[\ket{\Psi\left(t;1\right)}\bra{\Psi\left(t;1\right)}\right].
    \label{eq:time_scaled}
\end{align}
 This relation justifies the notations of Eq.~\eqref{def:psi_t}. 
By combining Eq.~\eqref{eq:scaled_cMPS_def} with Eq.~\eqref{eq:Schrodinger_eq} and $t=\tau\theta$, we obtain
\begin{align}
    &\mathcal{H}(t)\ket{\Psi(t)}=\frac{i}{\tau}\partial_\theta \ket{\Psi(\tau;\theta)}=\frac{i}{\tau}\int_0^{\tau-ds}  \mathcal{V}\left(\tau, s+ds; \theta\right)d\mathcal{V}\left(s; \theta\right)\ket{\Psi\left(s\theta\right)},
    \label{eq:H_rep}
\end{align}
where 
\begin{align}
    \label{eq:dV}
    d\mathcal{V}\left(s; \theta\right)&\equiv -iH_{\mathrm{eff}}\otimes \mathbb{I}_F ds+\frac{1}{2\sqrt{\theta}}\sum_m L_m \otimes d\phi_m^\dagger(s), \\
    d\phi_m^\dagger(s)&\equiv \int_s^{s+ds} ds' \phi_m^\dagger(s').    
\end{align}
 In Eq.~\eqref{eq:H_rep}, note that the integral range is $[0, \tau-ds]$ because of the term $\mathcal{V}\left(\tau, s+ds; \theta\right)$. 
We can verify that the variance of the induced Hamiltonian corresponds to the quantum Fisher information as follows.
The mean of the induced Hamiltonian is given by
\begin{align}
    \bra{\Psi(t)}\mathcal{H}(t)\ket{\Psi(t)}=i\braket{\Psi(t)|d_t\Psi(t)},
\end{align}
where $\ket{d_t\Psi(t)}\equiv d/dt\ket{\Psi(t)}$.
Similarly, the variance of the induced Hamiltonian is given by
\begin{align}
    &\bra{\Psi(t)}\mathcal{H}(t)^{2}\ket{\Psi(t)}-\bra{\Psi(t)}\mathcal{H}(t)\ket{\Psi(t)}^{2}\nonumber\\&=\braket{d_{t}\Psi(t)|d_{t}\Psi(t)}-|\braket{\Psi(t)|d_{t}\Psi(t)}|^{2}\nonumber\\&=\frac{1}{4}\mathcal{J}(t)=\frac{1}{4t^{2}}\mathcal{B}(t),
    \label{eq:Hamiltonian_QDA}
\end{align}
where $\mathcal{J}(t)$ denotes the quantum Fisher information, and $\mathcal{B}(t)\equiv t^2\mathcal{J}(t)$ denotes the quantum dynamical activity.
First, we consider $A= \mathcal{C}_F\otimes \mathbb{I}_{S}$ and $B=\mathcal{H}(t)$ in Eq.~\eqref{eq:robertson_UR_def}.
From Eq.~\eqref{eq:Schrodinger_eq}, Eq.~\eqref{eq:Hamiltonian_QDA}, and $\bra{\Psi(t)}\mathcal{C}_F\otimes \mathbb{I}_{S}\ket{\Psi(t)}=\mathrm{Tr}_{F,S}[\ket{\Psi(t)}\mathcal{C}_F\otimes \mathbb{I}_{S}\bra{\Psi(t)}]=\mathrm{Tr}_{F}[\mathcal{C}_F\mathrm{Tr}_S[\ket{\Psi(t)}\bra{\Psi(t)}]]=\mathrm{Tr}_F[\mathcal{C}_F\rho_F(t)]$, we obtain the thermodynamic uncertainty relation:
\begin{align}
    \frac{\llbracket\mathcal{C}_F\rrbracket_F(t)^2}{t^2 \left(\partial_t \langle \mathcal{C}_F\rangle_F(t)\right)^2}\geq \frac{1}{\mathcal{B}(t)},
\end{align}
where 
\begin{align}
    \langle \bullet \rangle_Z(t)&\equiv \mathrm{Tr}_Z[\bullet \rho_Z(t)],\\
    \llbracket \bullet \rrbracket_Z(t)&\equiv \sqrt{\langle \bullet^2 \rangle_Z(t) -  \langle \bullet \rangle_Z(t)^2}.
\end{align}
Here, a subscript $Z$ denotes either $F$ (field) or $S$ (primary system). $\mathrm{Tr}_Z$ is the trace operation with respect to the field ($Z=F$) or the primary system ($Z=S$).
$\langle \bullet \rangle_Z(t)$ and $\llbracket \bullet \rrbracket_Z(t)$ correspond to the mean and standard deviation, respectively, with respect to the field or the primary system at time $t$. 
Next, consider the case where $A=\ket{\Psi(0)}\bra{\Psi(0)}$ and $B=\mathcal{H}(t)$.
In the closed quantum system, we can derive the Mandelstam-Tamm bound by setting $A=\ket{\psi(0)}\bra{\psi(0)}$ and assigning $B$ to the Hamiltonian in the Robertson uncertainty relation, where $\ket{\psi(0)}$ is the initial state vector.
In a similar way, from
Eq.~\eqref{eq:Hamiltonian_QDA}, we obtain
\begin{align}
    \frac{1}{2}\int_0^t ds \frac{\sqrt{\mathcal{B}(s)}}{s} \geq \mathcal{L}_D(\ket{\Psi(t)}, \ket{\Psi(0)}).
\end{align}
Using the monotonicity of the fideltiy $\mathcal{L}_D(\ket{\Psi(t)}, \ket{\Psi(0)})\geq  \mathcal{L}_D(\rho_S(t), \rho_S(0))$, we obtain Eq.~\eqref{eq:QSL_open_quantum}.
\section{Maccone-Pati uncertainty relation\label{sec:MacconePatiUR}}
We extend the Maccone-Pati uncertainty relation [Eq.~\eqref{eq:Maccone_Pati_def}, Eq.~\eqref{eq:Maccone_Pati2_def}] to the open quantum system in this section. 
Consider the state vectors of the primary system $\ket{\psi_S}$ and $\ket{\overline{\psi}_S}$ that are mutually orthogonal $\braket{\overline{\psi}_S|\psi_S} = 0$.
We define the following notations:
\begin{align}
    \ket{\overline{\Psi}(t)}&\equiv U(t)\ket{\overline{\psi}_S}\otimes\ket{\mathrm{vac}}, \\
    \label{eq:def_chi_Z}
    \chi_Z(t)&\equiv \mathrm{Tr}_Z\left[\ket{\overline{\Psi}(t)}\bra{\Psi(t)}\right],\quad (Z=F \;\mbox{or} \; S).
\end{align}
The operator $\chi_{S}(t)$ characterises the properties of coherence and its time evolution is governed by the Lindblad equation. 
Since $U(t)$ is unitary, it follows that $\braket{\Psi(t)|\overline{\Psi}(t)}=0$ and the trace of $\chi_{S}(t)$ is zero.
Therefore, we can apply Eq.~\eqref{eq:Maccone_Pati_def} for $\ket{\Psi(t)}$ and $\ket{\overline{\Psi}(t)}$.
We set the induced Hamiltonian $\mathcal{H}(t)$ to $B$ in Eq.~\eqref{eq:Maccone_Pati_def}. 
From Eq.~\eqref{eq:H_rep}, we obtain
\begin{align}
   & \tau\bra{\Psi(t)}\mathcal{H}(t)\ket{\overline{\Psi}(t)}\nonumber\\
   &=i\mathrm{Tr}_{F,S}\left[\ket{\partial_{\theta} \overline{\Psi}\left(\tau;\theta\right)}\bra{\Psi\left(\tau;\theta\right)}\right]\nonumber\\
   & =i\int_0^{\tau-ds}  \mathrm{Tr}_{F,S}\left[\mathcal{U}\left(\tau, s+ds; \theta\right) d\mathcal{V}\left(s; \theta\right)\ket{\overline{\Psi}(s\theta)}\bra{\Psi(s\theta)}\mathcal{V}\left(s+ds, s; \theta\right)^\dagger\mathcal{U}\left(\tau, s+ds; \theta\right)^\dagger\right]\nonumber\\
    & =i\int_0^{\tau-ds}  \mathrm{Tr}_{F,S}\left[\mathcal{V}\left(s+ds, s; \theta\right)^\dagger d\mathcal{V}\left(s; \theta\right)\ket{\overline{\Psi}(s\theta)}\bra{\Psi(s\theta)}\right]\nonumber\\
    &= i\int_0^{\tau} ds \mathrm{Tr}_{S}\left[\left(-iH_{\mathrm{eff}}+\frac{1}{2}\sum_m L_m^\dagger L_m \right)\chi_S(s\theta)\right]=\frac{1}{\theta}\int_0^{t}ds  \mathrm{Tr}_{S}\left[H_S\chi_S(s)\right],
    \label{eq:H_braket}
\end{align}
where we use the unitarity of $\mathcal{U}$ in the third equality and  the canonical commutation relation of $\phi_m$ and $\phi_{m'}^{\dagger}$  in the forth equality. 
This relation yields
\begin{align}
    \label{eq:mean_H_chi}
    \bra{\Psi(t)}\mathcal{H}(t)\ket{\overline{\Psi}(t)}=\frac{1}{t}\int_0^{t}ds  \mathrm{Tr}_{S}\left[H_S\chi_S(s)\right].
\end{align}
Suppose that $\mathbb{I}_{S}\otimes \mathcal{C}_F$ and $\mathcal{C}_S\otimes \mathbb{I}_{F}$ are time-independent.
 Substituting these observables into $A$ 
in Eq.~\eqref{eq:Maccone_Pati_def}, we obtain
\begin{align}
    \label{eq:F_braket}
    \bra{\Psi(t)}\mathbb{I}_{S}\otimes \mathcal{C}_F\ket{\overline{\Psi}(t)}=\mathrm{Tr}_F[\mathcal{C}_F\chi_F(t)],\\
    \label{eq:S_braket}
    \bra{\Psi(t)}\mathcal{C}_S\otimes \mathbb{I}_{F}\ket{\overline{\Psi}(t)}=\mathrm{Tr}_S[\mathcal{C}_S\chi_S(t)],
\end{align}
where we used Eq.~\eqref{eq:def_chi_Z}.
Substituting Eq.~\eqref{eq:Hamiltonian_QDA}, Eq.~\eqref{eq:H_braket} and Eq.~\eqref{eq:F_braket} (or Eq.~\eqref{eq:S_braket}) into Eq.~\eqref{eq:Maccone_Pati_def} and Eq.~\eqref{eq:Maccone_Pati2_def}, we derive the following inequalities.
\begin{align}
    \label{eq:open_Maccone-Pati_1}
    &\frac{\mathcal{B}(t)}{4}+t^{2}\llbracket\mathcal{C}_{Z}\rrbracket_{Z}^{2}\geq\pm t^{2}\partial_{t}\langle\mathcal{C}_{Z}\rangle_{Z}+\left|\int_{0}^{t}ds\mathrm{Tr}_{S}\left[H_{S}\chi_{S}(s)\right]\pm it\mathrm{Tr}_{Z}[\mathcal{C}_{Z}\chi_{Z}(t)]\right|^{2}, \\
    \label{eq:open_Maccone-Pati_2}
    &\sqrt{\mathcal{B}(t)}\dblbrace{\mathcal{C}_{Z}}_{Z}\geq\frac{\pm t\partial_{t}\braket{\mathcal{C}_{Z}}_{Z}}{1-\frac{1}{2}\left|\frac{2\int_{0}^{t}ds\mathrm{Tr}_{S}\left[H_{S}\chi_{S}(s)\right]}{\sqrt{\mathcal{B}(t)}}\pm i\frac{\mathrm{Tr}_{Z}[\mathcal{C}_{Z}\chi_{Z}(t)]}{\dblbrace{\mathcal{C}_{Z}}_{Z}}\right|^{2}}.
\end{align}
Here, we choose the sign such that the right-hand side is positive. We omit the arguments $t$ in $\langle\bullet\rangle_Z$ and $\llbracket\bullet\rrbracket_Z$.
\subsection{Number operator case}
Let us consider a  generalised  number operator: 
\begin{align}
    \mathcal{C}_{F}\equiv\sum_{n_{1},n_{2},\cdots n_{N_{C}}}\left(\sum_{m}\alpha_{m}n_{m}\right)\Pi_{n_{1}n_{2}\cdots n_{N_{C}}},
    \label{eq:def_number_op}
\end{align}
where the eigenvalue $n_m$ denotes the number of  the  $m$th jumps within $[0,\tau]$ 
and $\Pi_{n_{1}n_{2}\cdots n_{N_{C}}}$ is a corresponding projector to the space of $\{n_{1}, n_{2}, \cdots, n_{N_{C}}\}$.  The coefficients $\{\alpha_m\}$ are weights associated to the physical process one dealing with (e.g. $\alpha_m=\pm 1$).  
Note that, using the total number operator $\int_0^\tau \phi_m^\dagger(s)\phi_m(s)ds$, $\mathcal{C}_F$ can be expressed by
\begin{align}
    \mathcal{C}_{F}=\sum_{m}\alpha_{m}\int_0^\tau\phi_{m}^{\dagger}(s)\phi_{m}(s)ds.
    \label{eq:CF_by_total_number_op}
\end{align}
We introduce the characteristic function to obtain moments of $\mathcal{C}_F$ and derive its master equation in this section.
The characteristic function is defined as 
\begin{align}
    \label{eq:def_varphi_S}
    \varphi_{S}(\tau;\theta, \xi)\equiv \mathrm{Tr}_F\left[e^{i\xi\mathcal{C}_{F}}\ket{\Phi(\tau;\theta)}\bra{\Psi(\tau;\theta)}\right],
\end{align}
where $\ket{\Phi(\tau;\theta)}$ represents $\ket{\Psi(\tau;\theta)}$ or $\ket{\overline{\Psi}(\tau;\theta)}$. From Eq.~\eqref{eq:def_number_op}, we have
\begin{align}
    e^{i\xi\mathcal{C}_{F}}&=\sum_{n_{1},n_{2},\cdots n_{N_{C}}}e^{i\xi\sum_{m}\alpha_{m}n_{m}}\Pi_{n_{1}n_{2}\cdots n_{N_{C}}},\\
    \label{eq:varphi_decomposition}
     \mathrm{Tr}_F\left[e^{i\xi\mathcal{C}_{F}}\ket{\Phi(\tau;\theta)}\bra{\Psi(\tau;\theta)}\right]&=\sum_{n_{1},n_{2},\cdots n_{N_{C}}}e^{i\xi\sum_{m}\alpha_{m}n_{m}}\varphi_{S;\;n_{1},n_{2},\cdots n_{N_{C}}}(\tau;\theta),
\end{align}
where $\varphi_{S;\;n_{1},n_{2},\cdots n_{N_{C}}}(\tau;\theta)$ is the operator given the number of jumps as $\{n_{1}, n_{2}, \cdots, n_{N_{C}}\}$, and the original operator can be recovered as $\varphi_{S}(\tau;\theta, \xi=0)=\sum_{n_{1},n_{2},\cdots n_{N_{C}}}\varphi_{S;\;n_{1},n_{2},\cdots n_{N_{C}}}(\tau;\theta)$. Following the same procedure used to derive Eq.~\eqref{eq:time_scaled}, we obtain
\begin{align}
    \varphi_{S}(\tau;\theta, \xi)=\varphi_{S}(t;1, \xi)\equiv\varphi_{S}(t; \xi) .
    \label{eq:time_scaled2}
\end{align}
  Note that the time interval of $\mathcal{C}_F$ is $[0,t]$ for $\varphi_{S}(t;1, \xi)$. 
From Eq.~\eqref{eq:varphi_decomposition} and using the notation $\ket{\Phi(t)}=\ket{\Phi\left(\tau;\theta\right)}$, we obtain $\mathrm{Tr}_F\left[e^{i\xi\mathcal{C}_{F}}d\phi_m^\dagger(t)\ket{\Phi(t)}\bra{\Psi(t)}d\phi_m(t)\right]=e^{i\xi\alpha_m}\mathrm{Tr}_F\left[e^{i\xi\mathcal{C}_{F}}\ket{\Phi(t)}\bra{\Psi(t)}\right]dt$. By combining this relation with Eq.~\eqref{eq:def_V} and Eq.~\eqref{eq:time_scaled2}, it can be verified that $\varphi_{S}(t; \xi)$ obeys the tilted Lindblad equaion~\cite{Landi:2023:CurFlucReview}:
\begin{align}
    \label{eq:tilted_Lindblad}
    \partial_t \varphi_{S}(t; \xi)&=\mathcal{L}[\xi]\varphi_{S}(t; \xi),\\
    \mathcal{L}[\xi]\bullet&\equiv -i[H_S,\bullet]+\sum_m \left(e^{i\xi\alpha_m}L_m \bullet L_m^\dagger -\frac{1}{2} \{L_m^\dagger L_m, \bullet\}\right).
\end{align}
Letting $\varphi_{F}(t)\equiv \mathrm{Tr}_S\left[\ket{\Phi(t)}\bra{\Psi(t)}\right]$, we can obtain moments of $\mathcal{C}_F$ with respect to the field by differentiating the solution of the tilted Lindblad equation $\varphi_{S}(t;\xi)$:
\begin{align}
    \mathrm{Tr}_{F}\left[\mathcal{C}_F^k\varphi_{F}(t)\right]=\braket{\Psi(t)|\mathcal{C}_{F}^k|\Phi(t)}=(-i\left.\partial_{\xi})^k\braket{\Psi(t)|e^{i\xi\mathcal{C}_{F}}|\Phi(t)}\right|_{\xi=0}=(-i\left.\partial_\xi)^k\mathrm{Tr}_{S}\left[\varphi_{S}(t;\xi)\right]\right|_{\xi=0},
\end{align}
where we use Eq.~\eqref{eq:def_varphi_S}.
When $k=1$, by differentiating this equation with respect to $t$ and using Eq.~\eqref{eq:tilted_Lindblad}, we obtain 
\begin{align}
    \partial_{t}\mathrm{Tr}_{F}\left[\mathcal{C}_{F}\varphi_{F}(t)\right]&=-i\partial_{t}\left.\partial_{\xi}\mathrm{Tr}_{S}\left[\varphi_{S}(t;\xi)\right]\right|_{\xi=0}\nonumber\\&=-i\mathrm{Tr}_{S}\left[\left.\partial_{\xi}\left(\mathcal{L}[\xi]\varphi_{S}(t;\xi)\right)\right|_{\xi=0}\right]=\sum_{m}\alpha_{m}\mathrm{Tr}_{S}\left[L_{m}\varphi_{S}(t)L_{m}^{\dagger}\right],
\end{align}
where we use $\mathrm{Tr}_S[\mathcal{L}[0]\bullet]=\mathrm{Tr}_S[\mathcal{L}\bullet]=0$ and $\varphi_S(t)=\varphi_S(t;0)$.
Therefore, we obtain
\begin{align}
    \mathrm{Tr}_F\left[\mathcal{C}_F\varphi_{F}(t)\right]=\int_0^t ds \sum_m\alpha_m\mathrm{Tr}_S\left[ L_m\varphi_{S}(s)L_m^\dagger\right],
\end{align}
where we use $\mathrm{Tr}_F\left[\mathcal{C}_F\varphi_{F}(0)\right]=0$ since $\mathcal{C}_F \ket{\Phi(0)}=0$.  
Setting $\varphi=\rho$ or $\varphi=\chi$, we obtain the mean of $\mathcal{C}_F$ in Eq.~\eqref{eq:open_Maccone-Pati_1} and Eq.~\eqref{eq:open_Maccone-Pati_2}. 
\subsection{Quantum speed limit}
\label{Maccone_QSL}
We derive the QSL from  Eq.~\eqref{eq:Maccone_Pati_def}  by setting $A=\ket{\Psi(0)}\bra{\Psi(0)}$ and $B=\mathcal{H}(t)$. 
Consider the function $\varphi_S(\tau; \theta_1, \theta_2)\equiv \mathrm{Tr}_F\left[\ket{\Phi(\tau;\theta_1)}\bra{\Psi(\tau;\theta_2)}\right]$. 
This function obeys the two-sided Lindblad equation:
\begin{align}
    \label{eq:two_sided_lindblad}
     &\partial_\tau \varphi_S(\tau; \theta_1, \theta_2)=\mathcal{L}[\theta_1, \theta_2]\varphi_S(\tau; \theta_1, \theta_2),\\
     &\mathcal{L}[\theta_1, \theta_2]\bullet\equiv -i\theta_1 H_S \bullet + i\theta_2 \bullet H_S+ \sqrt{\theta_1\theta_2} L_m \bullet L_m^\dagger -\frac{1}{2}\sum_m\left( \theta_1 L_m^\dagger L_m \bullet +\theta_2 \bullet L_m^\dagger L_m \right).
\end{align}
Taking $\theta_1=\theta$ and $\theta_2=0$, we obtain 
\begin{align}
     &\partial_\tau \varphi_S(\tau; \theta, 0)=
       -i\theta  H_S \varphi_S(\tau; \theta, 0)- \frac{\theta}{2}\sum_m L_m^\dagger L_m  \varphi_S(\tau; \theta, 0)=-i\theta H_{\mathrm{eff}}\varphi_S(\tau; \theta, 0).
\end{align}
From this equation and $\varphi_S(0)=\varphi(0; \theta_1, \theta_2)$, it follows that $\varphi_S(\tau; \theta, 0)=\exp(-i H_{\mathrm{eff}} \theta\tau)\varphi_S(0)=\exp(-i H_{\mathrm{eff}} t)\varphi_S(0)$. By combining this equation with $\braket{\Psi(0)|\Phi(t)}=\mathrm{Tr}_{F,S}[\ket{\Phi(\tau;\theta)}\bra{\Psi(\tau;0)}]=\mathrm{Tr}_S[\varphi_S(\tau; \theta, 0)]$ and taking $\varphi=\rho$ or $\varphi=\chi$, we obtain
\begin{align}
    \label{eq:Psi_psi}
    &\braket{\Psi(t)|\Psi(0)}=\mathrm{Tr}_S[\exp(i H_{\mathrm{eff}}^\dagger t)\rho_S(0)]=\braket{\psi_S|\exp(i H_{\mathrm{eff}}^\dagger t)|\psi_S}\equiv \gamma(t), \\
    \label{eq:Psi_chi}
    &\braket{\Psi(0)|\overline\Psi(t)}=\mathrm{Tr}_S[\exp(-i H_{\mathrm{eff}}t)\chi_S(0)]=\braket{\psi_S|\exp(-i H_{\mathrm{eff}}t)|\overline\psi_S}.
\end{align}
From $\braket{\Psi(t)|A|\Psi(t)} = |\braket{\Psi(0)|\Psi(t)}|^2= |\gamma(t)|^2$ and 
$\dblbrace{A}=\sqrt{\braket{\Psi(t)|A^2|\Psi(t)}-\braket{\Psi(t)|A|\Psi(t)}^2}=\sqrt{\braket{\Psi(t)|A|\Psi(t)}-|\gamma(t)|^{4}}=\sqrt{|\gamma(t)|^{2}(1-|\gamma(t)|^{2})}$, we have
\begin{align}
    \label{eq:diff_gamma}
    \frac{i\braket{[A,\mathcal{H}]}}{2\dblbrace{A}}&=-\frac{d\braket{A}}{dt}\frac{1}{2\dblbrace{A}}=-\frac{\frac{d}{dt}\left(|\gamma(t)|^{2}\right)}{2\sqrt{|\gamma(t)|^{2}(1-|\gamma(t)|^{2})}}\nonumber\\&=-\frac{1}{\sqrt{1-|\gamma(t)|^{2}}}\frac{d}{dt}|\gamma(t)|=\frac{d}{dt}\arccos\left(|\gamma(t)|\right).
\end{align}
Substituting $A / \dblbrace{A}$ and $\mathcal{H} / \dblbrace{\mathcal{H}}$ into $A$ and $B$, respectively, in Eq.~\eqref{eq:Maccone_Pati_def}, we have 
\begin{align}
    2=\left\llbracket \frac{A}{\dblbrace{A}}\right\rrbracket ^{2}+\left\llbracket \frac{\mathcal{H}}{\dblbrace{\mathcal{H}}}\right\rrbracket ^{2}\geq\pm i\frac{\Braket{\left[A,\mathcal{H}\right]}}{\dblbrace{A}\dblbrace{\mathcal{H}}}+\left|\Braket{\psi\left|\frac{A}{\dblbrace{A}}\pm i\frac{\mathcal{H}}{\dblbrace{\mathcal{H}}}\right|\overline{\psi}}\right|^{2}.
    \label{eq:nonnegative_relation2}
\end{align} 
The Maccone-Pati uncertainty relation in Eq.~\eqref{eq:Maccone_Pati_def} is valid for both positive and negative signs. Threfore, we select the plus sign so that the sign of the first term in the right hand side in Eq.~\eqref{eq:nonnegative_relation2} corresponds to the sign of Eq.~\eqref{eq:diff_gamma}. Substituting Eqs.\eqref{eq:Hamiltonian_QDA}, \eqref{eq:mean_H_chi}, and \eqref{eq:Psi_psi}--\eqref{eq:diff_gamma} into Eq.~\eqref{eq:nonnegative_relation2}, we obtain
\begin{align}
    (1-R(t))\frac{\sqrt{\mathcal{B}(t)}}{2t}\geq \frac{d}{dt}\arccos\left(|\gamma(t)|\right),  \label{eq:R_inequality}
\end{align}
where 
\begin{align}
    R(t)&\equiv\frac{1}{2}\left|\Braket{\psi\left|\frac{A}{\dblbrace{A}} + i\frac{\mathcal{H}}{\dblbrace{\mathcal{H}}}\right|\overline{\psi}}\right|^{2}=\frac{1}{2}\left|\frac{\gamma(t)\mathrm{Tr}_S[\exp(-i H_{\mathrm{eff}}t)\chi_S(0)]}{|\gamma(t)|\sqrt{(1-|\gamma(t)|^2)}} + i\frac{2\int_{0}^{t}ds\mathrm{Tr}_{S}\left[H_{S}\chi_{S}(s)\right]}{\sqrt{\mathcal{B}(t)}}\right|^2.
    \label{eq:def_R}
\end{align}
Integrating Eq.~\eqref{eq:R_inequality} from $t=0$ to $\tau$, we obtain
\begin{align}
    \label{eq:QSL_Maccone}
    &\frac{1}{2}\int_0^\tau dt  (1-R(t))\frac{\sqrt{\mathcal{B}(t)}}{t}\geq \arccos(|\gamma(\tau)|)=\mathcal{L}_D(\ket{\Psi(\tau)}, \ket{\Psi(0)}) \geq  \mathcal{L}_D(\rho_S(\tau), \rho_S(0)),
\end{align}
where we use the monotonicity of the fidelity.

\section{Robertson-Schr\"odinger uncertainty relation\label{sec:Appendix_RSUR}}
We extend the Robertson-Schr\"odinger uncertainty relation [Eq.~\eqref{eq:Robertson_schrodinger_def}] to the open quantum system in this section. 
\subsection{Primary system operator case}
\label{sec:RSUR_TUR_system}
From Eq.~\eqref{eq:H_rep} and using the cyclic property of the trace, we obtain
\begin{align}
    &\tau\bra{\Psi(t)}\mathcal{C}_S\mathcal{H
}(t)\ket{\Psi(t)}=i\mathrm{Tr}_{F,S}\left[\mathcal{C}_S\ket{\partial_{\theta} \Psi\left(\tau;\theta\right)}\bra{\Psi\left(\tau;\theta\right)}\right]\nonumber\\
    & =i\int_0^{\tau-ds}  \mathrm{Tr}_{F,S}\left[\mathcal{C}_S \mathcal{V} \left(\tau, s+ds; \theta\right) d\mathcal{V}\left(s; \theta\right)\ket{\Psi(s\theta)}\bra{\Psi(s\theta)}\mathcal{V}\left(s+ds, s; \theta\right)^\dagger \mathcal{V}\left(\tau, s+ds; \theta\right)^\dagger\right]\nonumber\\
     &=i\int_0^{\tau-ds}  \mathrm{Tr}_{F,S}\left[ \mathcal{V}\left(s+ds, s; \theta\right)^\dagger\mathcal{V}\left(\tau, s+ds; \theta\right)^\dagger\mathcal{C}_S \mathcal{V} \left(\tau, s+ds; \theta\right) d\mathcal{V}\left(s; \theta\right)\ket{\Psi(s\theta)}\bra{\Psi(s\theta)}\right].    
     \label{eq:CH}
\end{align}
Substituting Eq.~\eqref{eq:def_V} and Eq.~\eqref{eq:dV} into this equation, and using  the canonical commutation relation of $\phi_m$ and $\phi_{m'}^{\dagger}$  for the interval $[s, s+ds)$, we obtain
\begin{align}
    \label{eq:CH2}
    &i\int_{0}^{\tau}ds\mathrm{Tr}_{F,S}\Biggl[\left(-i\mathcal{V}\left(\tau,s+ds;\theta\right)^{\dagger}\mathcal{C}_{S}\mathcal{V}\left(\tau,s+ds;\theta\right)H_{\mathrm{eff}}+\frac{1}{2}\sum_{m}L_{m}^{\dagger}\mathcal{V}\left(\tau,s+ds;\theta\right)^{\dagger}\mathcal{C}_{S}\mathcal{V}\left(\tau,s+ds;\theta\right)L_{m}\right)\nonumber\\&\times\ket{\Psi(s\theta)}\bra{\Psi(s\theta)}\Biggr]\nonumber\\
    &=\int_{0}^{\tau}ds\mathrm{Tr}_{S}\left[\left(\check{\mathcal{C}}_{S}\left(\theta(\tau-s)\right)H_{\mathrm{eff}}+\frac{i}{2}\sum_{m}L_{m}^{\dagger}\check{\mathcal{C}}_{S}\left(\theta(\tau-s)\right)L_{m}\right)\rho_{S}(\theta s)\right]\nonumber\\&=\frac{1}{\theta}\int_{0}^{t}ds\mathrm{Tr}_{S}\left[\left(\check{\mathcal{C}}_{S}(t-s)H_{\mathrm{eff}}+\frac{i}{2}\sum_{m}L_{m}^{\dagger}\check{\mathcal{C}}_{S}(t-s)L_{m}\right)\rho_{S}(s)\right],
\end{align}
where $\check{\mathcal{C}_S}(t)\equiv e^{\mathcal{L}^\dagger t}C_S$ is the Heisenberg interpretation of $\mathcal{C}_S$. 
Since $\bra{\Psi(t)}\{\mathcal{C}_S,\mathcal{H}(t)\}\ket{\Psi(t)}=2\mathrm{Re}\bra{\Psi(t)}\mathcal{C}_S\mathcal{H}(t)\ket{\Psi(t)}$, by combining Eq.~\eqref{eq:CH} with Eq.~\eqref{eq:CH2}, we obtain
\begin{align}
    \bra{\Psi(t)}\{\mathcal{C}_S,\mathcal{H}(t)\}\ket{\Psi(t)}=\frac{2}{t}\mathrm{Re}\int_0^{t}ds  \mathrm{Tr}_{S}\left[\check{\mathcal{C}}_S(t-s)H_{\mathrm{eff}}\rho_S(s)\right],
    \label{eq:anticom}
\end{align}
where we used the hermicity of $\check{\mathcal{C}}_{S}(s)$.
By substituting $\mathcal{C}_{S}=\mathbb{I}_S$ into Eq.~\eqref{eq:anticom}, we obtain
\begin{align}
    \label{eq:H_mean}
    \bra{\Psi(t)}\mathcal{H}(t)\ket{\Psi(t)}=\frac{1}{t}\int_0^{t}ds  \mathrm{Tr}_{S}\left[H_S\rho_S(s)\right].
\end{align}
Setting $A=\mathbb{I}_{F}\otimes \mathcal{C}_S$ and $B=\mathcal{H}(t)$ in Eq.~\eqref{eq:Robertson_schrodinger_def}, and combining these relations, we obtain  Eq.~\eqref{eq:new_rsr_main}. 
\subsection{Number operator case}
\label{sec:RSUR_TUR_field}
When $\mathcal{C}_F$ is given by Eq.~\eqref{eq:def_number_op}, we set  $A=\mathcal{C}_F\otimes \mathbb{I}_{S}$ and $B=\mathcal{H}(t)$ in Eq.~\eqref{eq:Robertson_schrodinger_def}. Let us define the characteristic function of $\mathcal{C}_F$ as
\begin{align}
    \label{eq:def_rho_character}
    \varphi_S(\tau; \theta_1, \theta_2,\xi)\equiv \mathrm{Tr}_F\left[e^{i\xi\mathcal{C}_{F}}\ket{\Psi(\tau;\theta_1)}\bra{\Psi(\tau;\theta_2)}\right].
\end{align}
In a similar manner to obtain Eq.~\eqref{eq:tilted_Lindblad}, we find that this function obeys the two-sided tilted Lindblad equation:
\begin{align}
    \label{eq:two_sided_tilted_lindblad}
     \partial_{\tau}\varphi_{S}(\tau;\theta_{1},\theta_{2},\xi)&=\mathcal{L}[\theta_{1},\theta_{2},\xi]\varphi_{S}(\tau;\theta_{1},\theta_{2},\xi),\\\mathcal{L}[\theta_{1},\theta_{2},\xi]\bullet&\equiv-i\theta_{1}H_{S}\bullet+i\theta_{2}\bullet H_{S}\nonumber\\&+\sum_{m}e^{i\xi\alpha_{m}}\sqrt{\theta_{1}\theta_{2}}L_{m}\bullet L_{m}^{\dagger}-\frac{1}{2}\sum_{m}\left(\theta_{1}L_{m}^{\dagger}L_{m}\bullet+\theta_{2}\bullet L_{m}^{\dagger}L_{m}\right).
\end{align}
From Eq.~\eqref{eq:def_rho_character}, we obtain
\begin{align}
     &\partial_\tau \mathrm{Tr}_F\left[\mathcal{C}_{F}\mathcal{H}(t)\ket{\Psi(t)}\bra{\Psi(t)}\right]=\left.\partial_\tau\partial_\xi \frac{1}{\tau}\partial_{\theta_1} \varphi_{S}(\tau; \theta_1, \theta, \xi)\right|_{\theta_1=\theta, \; \xi=0}\nonumber\\
     &=-\frac{1}{\tau}\mathrm{Tr}_F\left[\mathcal{C}_{F}\mathcal{H}(t)\ket{\Psi(t)}\bra{\Psi(t)}\right]+\frac{1}{\tau}\left.\partial_\xi \partial_{\theta_1} \partial_\tau \varphi_{S}(\tau; \theta_1, \theta, \xi)\right|_{\theta_1=\theta, \; \xi=0}.
     \label{eq:diff_tau_char}
\end{align}
Note that $\theta$ and $\tau$ are independent variables and $t=\theta\tau$.
By using Eq.~\eqref{eq:two_sided_tilted_lindblad}, the second term can be written as
\begin{align}
    \label{eq:diff_tau_char_second}
&\frac{1}{\tau}\left.\partial_\xi \partial_{\theta_1} \partial_\tau \varphi_{S}(\tau; \theta_1, \theta, \xi)\right|_{\theta_1=\theta, \; \xi=0}=\frac{1}{\tau}\partial_{\theta_{1}}\partial_{\xi}\mathcal{L}[\theta_{1},\theta,0]|_{\theta_{1}=\theta}\rho_{S}(t)-i\partial_{\xi}\mathcal{L}[\theta,\theta,0]\mathrm{Tr}_{F}\left[\mathcal{H}(t)\ket{\Psi(t)}\bra{\Psi(t)}\right]\nonumber\\&+\frac{i}{\tau}\partial_{\theta_{1}}\mathcal{L}[\theta_{1},\theta,0]|_{\theta_{1}=\theta}\mathrm{Tr}_{F}\left[C_{F}\ket{\Psi(t)}\bra{\Psi(t)}\right]+\theta\mathcal{L}\mathrm{Tr}_{F}\left[\mathcal{C}_{F}\mathcal{H}(t)\ket{\Psi(t)}\bra{\Psi(t)}\right]\nonumber\\&=\frac{i}{2\tau}\sum_{m}\alpha_{m}L_{m}\rho_{S}(t)L_{m}^{\dagger}+\theta\sum_{m}\alpha_{m}L_{m}\mathrm{Tr}_{F}\left[\mathcal{H}(t)\ket{\Psi(t)}\bra{\Psi(t)}\right]L_{m}^{\dagger}+\frac{i}{\tau}\mathcal{L}_{\theta_{1}}\mathrm{Tr}_{F}\left[\mathcal{C}_{F}\ket{\Psi(t)}\bra{\Psi(t)}\right]\nonumber\\&+\theta\mathcal{L}\mathrm{Tr}_{F}\left[\mathcal{C}_{F}\mathcal{H}(t)\ket{\Psi(t)}\bra{\Psi(t)}\right],
\end{align}
where 
\begin{align}
    \mathcal{L}_{\theta_1}\bullet\equiv
    \left(-i H_S\bullet +\frac{1}{2}\sum_m L_m \bullet L_m^\dagger -\frac{1}{2} \sum_m L_m^\dagger L_m \bullet\right).
\end{align}
Since $\mathrm{Tr}_S[\mathcal{L}_{\theta_1}\bullet]=-i\mathrm{Tr}_S[H_S\bullet]$ and $\mathrm{Tr}_S[\mathcal{L}\bullet]=0$, by taking the trace of Eq.~\eqref{eq:diff_tau_char} and using Eq.~\eqref{eq:diff_tau_char_second}, we obtain
\begin{align}
    &\partial_\tau \bra{\Psi(t)}\mathcal{C}_{F}\mathcal{H}(t)\ket{\Psi(t)}=-\frac{1}{\tau}\bra{\Psi(t)}\mathcal{C}_{F}\mathcal{H}(t)\ket{\Psi(t)}+\frac{i}{2\tau}\sum_m \alpha_m \mathrm{Tr}_S\left[L_m\rho_S(t)L_m^\dagger\right]+\nonumber\\
    &\theta\sum_m \alpha_m \bra{\Psi(t)}L_m^\dagger L_m \mathcal{H}(t)\ket{\Psi(t)}+\frac{1}{\tau}\bra{\Psi(t)}\mathcal{C}_FH_{S}\ket{\Psi(t)}.
\end{align}
Taking the real part of this equation, we obtain
\begin{align}
    \partial_t\left(t\bra{\Psi(t)}\{\mathcal{C}_F,\mathcal{H}(t)\}\ket{\Psi(t)}\right)=2\bra{\Psi(t)}\mathcal{C}_FH_{S}\ket{\Psi(t)}+t\sum_m \alpha_m \bra{\Psi(t)}\{L_m^\dagger L_m,\mathcal{H}(t)\}\ket{\Psi(t)},
\end{align}
where we used $t=\theta \tau$.
Integrating this relation with respect to $t$ and using Eq.~\eqref{eq:anticom}, we obtain
\begin{align}
   &\bra{\Psi(t)}\{\mathcal{C}_{F},\mathcal{H}(t)\}\ket{\Psi(t)}\nonumber\\&=\frac{2}{t}\int_{0}^{t}ds\bra{\Psi(s)}\mathcal{C}_{F}H_{S}\ket{\Psi(s)}+\frac{1}{t}\int_{0}^{t}dss\sum_{m}\alpha_{m}\bra{\Psi(s)}\{L_{m}^{\dagger}L_{m},\mathcal{H}(s)\}\ket{\Psi(s)}\nonumber\\&=\frac{2}{t}\int_{0}^{t}ds\bra{\Psi(s)}\mathcal{C}_{F}H_{S}\ket{\Psi(s)}+\frac{2}{t}\mathrm{Re}\int_{0}^{t}ds_{1}\int_{0}^{s_{1}}ds_{2}\mathrm{Tr}_{S}\left[\check{\mathfrak{L}}(s_{1}-s_{2})H_{\mathrm{eff}}\rho_{S}(s_{2})\right],
\end{align}
where $\mathfrak{L}\equiv \sum_m \alpha_m L_m^\dagger L_m$.
By combining Eq.~\eqref{eq:Robertson_schrodinger_def} and Eq.~\eqref{eq:H_mean} with this equation, we obtain 
 Eq.~\eqref{eq:new_rsr_field_main}.
\subsection{Quantum speed limit\label{sec:RSUR_QSL}} 
We set $A=\ket{\Psi(0)}\bra{\Psi(0)}$ and $B=\mathcal{H}(t)$ in Eq.~\eqref{eq:Robertson_schrodinger_def}. The main difference from Section~\ref{Maccone_QSL} is the term $\braket{\{A,B\}}$. From Eq.~\eqref{eq:Psi_psi}, we obtain
\begin{align}
    \braket{\Psi(0)|\mathcal{H}|\Psi(t)}&=i\braket{\Psi(0)|d_{t}\Psi(t)}=\braket{\psi_{S}|H_{\mathrm{eff}}\exp(-iH_{\mathrm{eff}}t)|\psi_{S}}\nonumber\\&=\mathrm{Tr}_{S}[H_{\mathrm{eff}}\exp(-iH_{\mathrm{eff}}t)\rho_{S}(0)]\equiv\beta(t).
\end{align}
By combining this relation with Eq.~\eqref{eq:Psi_psi}, the term $\braket{\{A,B\}}$ yields
\begin{align}
    \braket{\{A,B\}}=2\mathrm{Re}\left(\gamma(t)\beta(t)\right).
\end{align}
From 
$\dblbrace{A}=\sqrt{|\gamma(t)|^{2}(1-|\gamma(t)|^{2})}$, which is shown in \ref{Maccone_QSL}, and using Eq.~\eqref{eq:diff_gamma}, we obtain
\begin{align}
     \sqrt{\frac{\mathcal{B}(t)}{4t^2}-S(t)}\geq \frac{d}{dt}\arccos\left(|\gamma(t)|\right),
\end{align}
where
\begin{align}
     S(t)\equiv \frac{1}{|\gamma(t)|^2(1-|\gamma(t)|^{2})}\left(\mathrm{Re}\left(\gamma(t)\beta(t)\right)-\frac{|\gamma(t)|^2}{t}\int_0^{t}ds  \mathrm{Tr}_{S}\left[H_S\rho_S(s)\right]\right)^2\geq 0.
\end{align}
Following the same procedure used to derive Eq.~\eqref{eq:QSL_Maccone}, we obtain  Eq.~\eqref{eq:RSUR_QSL}.

\section*{Acknowledgments}
This work was supported by JSPS KAKENHI Grant Numbers JP22H03659 and JP23K24915.

\providecommand{\newblock}{}

\end{document}